\documentclass{sig-alternate}
\usepackage[english]{babel}
\usepackage[latin1]{inputenc}
\usepackage{amssymb}
\usepackage[ruled]{algorithm2e}
\usepackage{bm}
\usepackage{cite}
\usepackage{amsmath}
\usepackage{graphicx}
\usepackage{epsfig}
\usepackage{url}
\usepackage{psfrag}
\usepackage[usenames]{color}
\usepackage{balance}
\usepackage{pstricks}
\usepackage{times}
\usepackage{array}
\usepackage{subfigure}
\usepackage{float}
\usepackage{multirow}

\newenvironment{pf}{{\it Proof. }}{\hfill $\square$\medskip}

\newcommand{\comment}[1]{}

\newtheorem{theorem}{Theorem}
\newtheorem{lemma}{Lemma}
\newcommand {\bpi}{\mbox{\boldmath $\pi$}}
\newcommand {\bphi}{\mbox{\boldmath $\phi$}}

\newcommand {\btau}{\mbox{\boldmath $\tau$}}
\newcommand {\cN}{\mathcal{N}}

\newcommand{\squishlist}{
 \begin{list}{$\bullet$}
  { \setlength{\itemsep}{0pt}
     \setlength{\parsep}{3pt}
     \setlength{\topsep}{3pt}
     \setlength{\partopsep}{0pt}
     \setlength{\leftmargin}{1.5em}
     \setlength{\labelwidth}{1em}
     \setlength{\labelsep}{0.5em} } }

\newcommand{\squishend}{
  \end{list}  }

\begin{document}

\title{Practical Characterization of Large Networks Using\\ Neighborhood Information}
\numberofauthors{1}
\author{
\alignauthor
     Pinghui Wang$^{1}$, Junzhou Zhao$^{2}$, Bruno Ribeiro$^{3}$, John C.S.\ Lui$^{1}$,\\ Don Towsley$^{3}$, and Xiaohong Guan$^{2}$\\
       \affaddr{$^{1}$Department of Computer Science and Engineering, The Chinese University of Hong Kong, Hong Kong}\\
       \affaddr{$^{2}$MOE Key Laboratory for Intelligent Networks and Network Security, Xi'an Jiaotong University, China}\\
       \affaddr{$^{3}$Department of Computer Science, University of Massachusetts Amherst, MA, USA}\\
       {\{phwang, jzzhao, xhguan\}@sei.xjtu.edu.cn, cslui@cse.cuhk.edu.hk,\\ \{towsley, ribeiro\}@cs.umass.edu}
}

\maketitle

\begin{abstract}
Characterizing large online social networks (OSNs)
through node querying is a challenging task.
OSNs often impose severe constraints on the query rate,
hence limiting the sample size to a small fraction of the total network.
Various ad-hoc subgraph sampling methods have been proposed,
but many of them give biased estimates
and no theoretical basis on the accuracy.
In this work, we focus on developing sampling methods for OSNs where querying
a node also reveals partial structural information about its neighbors.
Our methods are optimized for NoSQL graph databases
(if the database can be accessed directly),
or utilize Web API available on most major OSNs for graph sampling.
We show that our sampling method has provable convergence guarantees
on being an unbiased estimator,
and it is more accurate than current state-of-the-art methods.
We characterize metrics such as
node label density estimation and edge label density estimation,
two of the most fundamental network characteristics from which other network characteristics can be derived.
We evaluate our methods on-the-fly over several live networks using
their native APIs.
Our simulation studies over a variety of offline datasets show that
by including neighborhood information, our method drastically (4-fold) reduces the number of samples required
to achieve the same estimation accuracy of state-of-the-art methods.
\end{abstract}


\section{Introduction} \label{sec:introduction}

The literature on sampling large networks is vast and rich.
Various techniques have been proposed for subgraph sampling and characterization of
large networks~\cite{Leskovec2006,ICDM_Hubler08,KDD_Maiya11}
(refer to Ahmed et al.~\cite{AX_Neville12} for a good survey).
These techniques, however, often lack provable guarantees.
This means that after sampling a fraction of a large network,
one has no guarantees whether the metrics obtained are to be trusted.
Fortunately, researchers have recently made a push
towards network characterization through sampling with provable
properties and accuracy guarantees.

Techniques adapted to sample networks stored at NoSQL graph databases
or accessible from Web APIs
(e.g.\ available on Facebook, Foursquare, Pinterest, among others)
must refrain from randomly sampling too many nodes and all
together avoid sampling edges,
either due to caching inefficiencies or limitations in the API.
In practice, most online social networks (OSNs),
including those we present in this study,
do not provide random sampling primitives.
Practitioners perform random sampling by guessing user IDs in the user ID space,
which, if sparsely populated, imposes a large number of query misses
until a valid user is found.
In this context, techniques that heavily rely on random sampling,
such as Dasgupta et al.~\cite{Dasgupta2012},
suffers from low query rate.
Dasgupta et al.\ partially compensates the low query rate through
the use of neighborhood information present in the node query reply
of a number of major OSNs (e.g.\ Foursquare, Pinterest, Sina microblog).
Similarly, graph streaming techniques, such as
Ahmed et al.~\cite{AX_Neville12}, are also not well
adapted to this environment as
they require visiting all edges, which is prohibitively expensive in
a large network with millions or even billions of edges.

Recently, great focus has been placed on developing techniques that use
specially constructed ``{\em crawlers}'' to query the network
and to provide asymptotically unbiased estimates
of a handful of network characteristics~\cite{IMC_Ribeiro10,Gjoka2010}.
Chief among these techniques are random walks,
which provide provable accuracy and convergence guarantees
(see Ribeiro and Towsley\cite{CDC_Ribeiro12} and Avrechenkov et al.~\cite{AvrachenkovWAW2010}).
Random walks present a number of desirable properties that are
useful to characterize large networks;
(1) they require either few or no independently sampled nodes
    and produce asymptotically unbiased estimates and accuracy
    guarantees under mild conditions for a large family of directed\footnote{In directed networks where querying a node retrieves both the incoming and outgoing edges of that node.}
    and undirected networks, even when the network has multiple disconnected components, as long as some limited amount of random sampling is available~\cite{CDC_Ribeiro12,IMC_Ribeiro10,AvrachenkovWAW2010},
(2) use crawling to collect samples (which effectively implements importance
     sampling on node degrees), and
     can achieve relatively high query rates on NoSQL graph databases or
     using Web APIs, and
(3) does not require any advance knowledge of the network,
    such as its size or topology.
However, existing random walk (RW) techniques do not take advantage of
the extra neighborhood information,
despite the fact that neighborhood information is readily available in many OSNs at (practically) no sampling cost (obtained from the node query reply).
Including such extra information in RW-based estimator while
retaining unbiased guarantees is challenging due to different types of
biases involved in the sampling process.
\\

\noindent
{\bf Contributions:}
In this work, we consider the generalization of RW sampling
and combine current state-of-the-art estimators to include
neighborhood information.
Our estimator drastically reduces (by 4-fold) the number of samples required
to achieve the same estimation accuracy.
Examples of OSNs that provide neighborhood information are found
everywhere, e.g.\ Pinterest~\cite{Pinterest},
Foursquare~\cite{Foursquare},
Sina microblog~\cite{SinaMicroblog}, and Xiami~\cite{Xiami}.
Our generalization allows us
to include neighboring information in the estimation of a
variety of network characteristics from nodes sampled
using a random walk-based technique called Frontier Sampling~\cite{IMC_Ribeiro10}.
We also implement our method to sample networks in the wild,
and discover that the degree distribution of Foursquare
{\em does not} exhibit a heavy tail,
and, by using adapted versions of state-of-the-art algorithms we
also estimate that the average distance between users to be 5.8,
which is between values of average distances observed for
Twitter (4.1) and MSN messenger network (6.6)~\cite{LeskovecWWW2008,Kwak2010}.

This paper is organized as follows.
Several basic sampling techniques are summarized
in Section~\ref{sec:preliminaries}.
In Sections~\ref{sec:nodelabeldensity}, we present
the methodology of using neighborhoold information to estimate node label density.
In Section ~\ref{sec:edgelabeldensity},
we propose methods using neighborhood information
to estimate edge label density.
The performance evaluation and testing results are
presented in Section~\ref{sec:results}.
Section~\ref{sec:application} presents applications of
our methods on Foursquare and Pinterest websites.
Section~\ref{sec:related}
summarizes related work.
Section~\ref{sec:conclusions} concludes.


\section{Preliminaries} \label{sec:preliminaries}
In this section we introduce {\em Frontier Sampling} (FS), a generalization of random walk sampling methods~\cite{IMC_Ribeiro10}.
For ease of presentation, we assume undirected, connected, and non-bipartite networks.
Unless we state otherwise, denote by $G_d=(V, E_d)$ the directed graph under study,
and $G=(V, E)$ the undirected graph generated by ignoring the direction of edges in $G_d$.

The above assumptions are not too restrictive for the following reasons:
For {\em directed networks}, our method can be trivially adapted to include directed OSNs
such as Twitter and Flickr, which provide direct Web API access to the incoming and
outgoing edges of a node (such that our crawler can traverse on an undirected version
of the network).
Sometimes, there is a cost associated with obtaining the incoming and outgoing
edges of nodes with large in or out degrees
(e.g., Flickr provides only up to one hundred incoming or outgoing edges per query).
For {\em connected networks}, our previous work~\cite{AvrachenkovWAW2010}
consider random walk sampling and show how to augment
disconnected graphs using randomly sampled nodes into connected graphs
without changing the properties of the estimators.
Note that PageRank-style jumps are not suited for the task as they create unknown
biases in the estimators, see~\cite{AvrachenkovWAW2010}.
A trivial adaptation of the above argument can be used to show that
FS retains its properties on disconnected or
bipartite networks if a limited amount of random sampling is
available (e.g.\ one hundred sampled nodes in a network with millions of nodes).

Our accuracy guarantees follow directly from our results in
Ribeiro and Towsley~\cite{CDC_Ribeiro12}, which provides provable guarantees on the
mean squared error (MSE) accuracy of the degree distribution estimates
given by random walk sampling as a function of the number of samples and
the first nontrivial eigenvalue of the Laplacian of $G$.

\subsection {Frontier Sampling (FS)}
Frontier Sampling~\cite{IMC_Ribeiro10} is a fully distributed sampling algorithm that
performs $m$ independent RWs on $G$.
If $m=1$, FS behaves exactly like a RW.
When $m > 1$, compared to a single RW, FS can be more robust to the problems that arise from the walker getting trapped at a loosely connected component of $G$.
The $k$-th FS walker starts at node $s^{(k)}_0$, $k = 1,\ldots,m$.
Each FS walker has a  predefined budget $T$ (we explain how $T$ is chosen at the end of this section).
Denote by $\mathcal{N}(u)$ the set of neighbors of any node $u\in V$,
and by \mbox{$d_u=|\mathcal{N}(u)|$} the degree of $u$.
At each step an FS walker at node $u$  moves to a randomly node from $\cN(u)$, deducting from the budget $T$ a
random quantity $X \sim \mbox{Exp}(d_u)$, an exponentially distributed random variable with mean $1/d_u$.
FS stops when $T$ becomes negative.
If $G$ is a connected and non-bipartite graph, the probability that a node $v$ is sampled by FS converges to the following distribution
\[
\pi_v^{\text{FS}}=\frac{d_v}{2|E|}, \quad v\in V.
\]

FS can also be used to sample edges randomly, as the probability of traversing an edge $(u,v)\in E$ converges to
the uniform distribution~\cite{IMC_Ribeiro10}, that is
\begin{equation*}
\rho_{u,v}=\frac{1}{|E|}, \quad (u,v)\in E.
\end{equation*}
The choice of budget $T$ is often defined as the average number of nodes that one wishes to sample, $n$, divided by the number of FS walkers $m$ times the average degree, $\bar{d} $. In practice, one does not need to know $\bar{d}$ as $T$ may be increased dynamically on-the-fly.
Because we can adjust $T$ on-the-fly, in what follows we take the liberty to assume that FS samples exactly $n$ nodes.

We merge all $n$ samples collected by the FS walkers into a single stream $(s_1,\ldots,s_n)$ in any order.
Let $s_i$ be the $i$-th node sampled by FS, $i\ge 1$.
Let ``a.s.'' denote ``almost sure'' convergence, i.e., that the event of interest happens with probability one.
Then,
\begin{lemma}[Theorem 4.1~\cite{IMC_Ribeiro10}]\label{lemma:node}
For any function $\phi(v):V\rightarrow \mathbb{R}$, where $\sum_{\forall v\in V} \phi(v)<\infty$,
\[
\lim_{n\rightarrow \infty} \frac{1}{n} \sum_{i=1}^n \phi(s_i) \xrightarrow{a.s.} \sum_{\forall v\in V} \phi(v)\pi_v^{\text{\em FS}} \, .
\]
%
\end{lemma}
An important property of FS is that seeding $m \gg 1$ walkers with i.i.d.\ nodes sampled uniformly at random (UNI) is equivalent to initializing walkers in steady state~\cite[Theorem 5.4]{IMC_Ribeiro10}.
In practice, $m = 100$ initial UNI samples nodes, $(s^{(1)}_0,\ldots,s^{(100)}_0)$, are enough to initialize the FS walkers close to their joint steady state (thus, requiring only 100 ``expensive'' UNI sampled nodes).
This property of FS, sampling nodes according to their degree but being initialized in steady state using UNI,
is the result of a Markov chain trick used to uniformize continuous time Markov chains into discrete transition probability matrices.
For more details see Ribeiro and Towsley~\cite{IMC_Ribeiro10}.

In practice FS also works on disconnected graph as long as the initial choice of nodes is chosen from UNI and $m$ is large.
However, for disconnected graphs no convergence property can be shown as the different FS walkers do not mix.
In the absence of UNI samples, FS behaves  much like a single RW and it is advised that $m$ should be kept reasonably small.
Recent results in Ribeiro and Towsley~\cite{CDC_Ribeiro12} show provable guarantees of accuracy of RW-based methods:
\begin{lemma}[Theorem III.1~\cite{CDC_Ribeiro12}]\label{lemma:accuracy}
Let $w_1,\ldots,w_{n}$ be a set of nodes sampled independently proportionally to their degree, and $s_1,\ldots,s_{n}$ is a sequence of RW sampled nodes, then
\begin{equation}\label{eq:b}
\text{MSE}\left(\frac{1}{n} \sum_{i=1}^{n} \phi(s_i) \right) \leq \frac{1}{1-\alpha}  \text{MSE}\left(\frac{1}{n} \sum_{i=1}^{n} \phi(w_i) \right)\, , \: \forall k,
\end{equation}
where $\alpha$ is the first nontrivial eigenvalue of the Laplacian, $\mathcal{L}$, of $G$. The bound is tight~\cite{CDC_Ribeiro12}.
\end{lemma}
The bound in Lemma~\ref{lemma:accuracy} shows that the increase in MSE of RW (FS with $m=1$) sampling is at most $1/(1-\alpha)$ times larger than the MSE of independently sampling nodes according to their degree (importance sampling proportional to the degree).
The value of $\alpha$ goes to zero as the graph gets more connected, reducing the gap between MSEs of RW and i.i.d.\ importance sampling.

The value of $\alpha$ can also be ``artificially'' decreased using the RW with restarts (RWRST)~\cite{AvrachenkovWAW2010}, which augments the graph with edges of small weights.
RWRSTs are not to be confused with PageRank~\cite{PageRank} as the two Markov chains have remarkably different statistical properties.
Moreover, empirically FS achieves similar fast mixing if seed nodes are chosen uniformly (UNI) and $n \gg m$~\cite{IMC_Ribeiro10}.
Not surprisingly, FS behavior is remarkably similar to RWRST.
A RWRST is a RW that at node $v \in V$ chooses to jump to a randomly chosen node with probability $h / (d_v + h)$ or select a neighbor of $v$ with probability $d / (d_v + h)$, where $h>0$ is a parameter of the algorithm.
A RWRST stopped at the $m+1$ restart can be emulated by $m$ independent RWs that at node $v \in V$ stop with probability $h / (d_v + h)$.
Using results in Avrachenkov et al.~\cite[Theorem 2.1]{AvrachenkovWAW2010} it is easy to show that the MSE of a RWRST over a $d$-regular bounded by  multiplying the right hand side of~\eqref{eq:b} by $(1-\alpha)/(1 - \alpha d/(d+h))$.\footnote{Note that in a $d$-regular graph one should think that $\phi$ is a meaningful density function over the nodes, as estimating the degree distribution on a graph that only has degree $d$ is meaningless.}
Moreover, for any fixed number of sampled nodes $n$, the value of $h$ is a random variable that increases with $m$, implying that the latter MSE bound should decrease with $m$.
While this result is particular to $d$-regular graphs, these are likely to hold for a large class of graphs.
The similarity between FS and the $m$ simulated RWRST indicates that the FS MSE likely decreases with $m$, as long as $m \ll n$ and FS seed nodes are UNI sampled.
Unfortunately, a formal proof eluded us, as analyzing the FS Markov chain is more challenging than the analysis of RWRST in Avrachenkov et al.~\cite{AvrachenkovWAW2010} . We leave this analysis as future work.

\section{Node Label Density Estimation} \label{sec:nodelabeldensity}
In what follows we propose methods for estimating node label density.
Define $L(v)$ to be the node label of node $v$ under study, with range $\mbox{\boldmath $L$}=\{l_1,...,l_{K}\}$.
Denote by $\mbox{\boldmath $\theta$}=(\theta_1,\ldots,\theta_{K})$ the node label density, where $\theta_k$ ($1\le k \le K$) is the fraction of nodes with label $l_k$.
For example, when $L(v)$ is defined as the degree of node $v$,
then $\mbox{\boldmath $\theta$}$ is the node degree distribution of $G$.
If $L(v)$ denote the gender of node (or user) $v$,
then $\mbox{\boldmath $\theta$}$ is the gender distribution of the OSN under study.

\subsection{Simple Estimators of Node Densities}
To estimate $\mbox{\boldmath $\theta$}$ based on sampled nodes $\{s_i\}_{i=1,\ldots,n}$,
the stationary distribution of sampling methods (e.g. UNI, RW, and FS) $\bpi=(\pi_v: v\in V)$ is needed to correct the bias induced
by the underlying sampling method.
For $v\in V$, we have $\pi_v=\frac{1}{|V|}$ for UNI, and $\pi_v=\frac{d_v}{2|E|}$ for RW and FS.
Since the values of $|V|$ and $|E|$ are usually unknown,
unbiasing the error is not straightforward. Instead,
one may use a non-normalized stationary
distribution $\hat\bpi=(\hat\pi_v: v\in V)$ to reweight sampled
nodes $s_i$ ($1\le i\le n$), where $\hat\pi_v$ is computed as
\begin{equation}\label{eq:knownscale}
\hat\pi_v \propto \left\{
\begin{array}{ll}
  1 &\text{for UNI},\\
  d_v &\text{for RW and FS},\\
\end{array}
\right.
\end{equation}
Let $\mathbf{1}(\mathbf{P})$ be the indicator function that equals one when predicate $\mathbf{P}$ is true, and zero otherwise,
$\theta_k$ is estimated as follows
\begin{equation}\label{eq:firstestimatortheta}
\hat\theta_k=\frac{1}{C}\sum_{i=1}^n \frac{\mathbf{1}(L(s_i)=l_k)}{\hat\pi_{s_i}}, \quad 1\le k \le K,
\end{equation}
where $C=\sum_{i=1}^n {\hat\pi}_{s_i}^{-1}$.
Ribeiro and Towsley~\cite{IMC_Ribeiro10} shows that $\hat\theta_k$ ($1\le k \le K$) is an asymptotically unbiased estimate of $\theta_k$.

\subsection{Estimators Using Neighborhood Information of Sampled Nodes}
When the degrees and the node labels of sampled nodes' neighbors are available,
we propose the following estimator utilizing this free neighborhood information
\begin{equation}\label{eq:secondestimatortheta}
\breve\theta_k = \frac{1}{\breve C} \sum_{i=1}^n \sum_{w\in \mathcal{N}(s_i)} \frac{\mathbf{1}(L(w)=l_k)}{\hat\pi_{s_i} d_w}, \quad 1\le k \le K,
\end{equation}
where $\breve C=\sum_{i=1}^n \sum_{w\in \mathcal{N}(s_i)} \hat\pi_{s_i}^{-1} d_w^ {-1}$.
The above estimator is similar to one proposed in Dasgupta et al.~\cite{Dasgupta2012}.
However the estimator in Dasgupta et al.~\cite{Dasgupta2012} requires $|V|$ to be known in advance,
which is usually not available.
Moreover, Dasgupta et al.~\cite{Dasgupta2012} focuses on designing independent node
sampling methods (e.g. UNI, independent weighted node sampling),
which we argued has a low query rate.
Whereas we focus on crawling methods such as RW and FS.
For each node $v\in V$, Eq.~(\ref{eq:knownscale}) shows that $\pi_v/\hat\pi_v$ has the same value, denoted as $C_\pi$.
In what follows we analyze the accuracy of estimator $\breve\theta_k$ ($1\le k \le K$).

\begin{theorem}\label{theorem:nodelabel}
$\breve\theta_k$ ($1\le k \le K$) is an asymptotically unbiased  estimate of $\theta_k$.
\end{theorem}

\begin{pf} Applying Lemma~\ref{lemma:node}, we have
\begin{equation*}
\begin{split}
&\lim_{n\rightarrow \infty} \frac{1}{n} \sum_{i=1}^n  \left[\sum_{w\in \mathcal{N}(s_i)} \frac{\mathbf{1}(L(w)=l_k)}{\hat\pi_{s_i} d_w} \right]\\
& \xrightarrow{a.s.} \sum_{v\in V}\left(\pi_v \sum_{w\in \mathcal{N}(v)} \frac{\mathbf{1}(L(w)=l_k)}{\hat\pi_v d_w} \right)\\
&= C_\pi \sum_{w\in V}  \sum_{v\in \mathcal{N}(w)} \frac{\mathbf{1}(L(w)=l_k)}{d_w} \\
&= C_\pi \sum_{w\in V}  \mathbf{1}(L(w)=l_k)= C_\pi |V| \theta_k.
\end{split}
\end{equation*}
Similarly we prove that $\lim_{n\rightarrow \infty} \breve C/n \xrightarrow{a.s.} C_\pi |V| $. Therefore we have $\lim_{n\rightarrow \infty} \breve\theta_k \xrightarrow{a.s.} \theta_k$.
\end{pf}

We can easily find that neighbors of sampled nodes are biased to nodes with high degrees even for UNI.
Therefore, $\breve\theta_k$ is estimator based on biased samples.
Ribeiro and Towsley~\cite{IMC_Ribeiro10} shows that UNI has smaller mean square error (MSE) for small degree nodes than biased sampling methods such as RW.
It is consistent with our results in Section~\ref{sec:results}, which show that $\breve\theta_k$ may exhibit larger MSE than
$\hat\theta_k$ defined in~(\ref{eq:firstestimatortheta}).
Thus, we present the following mixture estimator for $\theta_k$
\begin{equation}\label{eq:mixestimator}
\hat\theta_k^{\text{mix}}= \alpha_k \hat\theta_k +(1-\alpha_k) \breve\theta_k, \quad 1\le k \le K,
\end{equation}
where parameter $\alpha_k$ lies between zero and one, and is used to determine the relative importance of two estimates $\hat\theta_k$ and $\breve\theta_k$.
Suppose that $\hat\theta_k$ and $\breve\theta_k$ are independent. Then $\hat\theta_k^{\text{mix}}$ has the smallest variance when $\alpha_k=\frac{\text{Var}(\breve\theta_k)}{\text{Var}(\hat\theta_k)+\text{Var}(\breve\theta_k)}$.

In what follows we propose estimators of node label density $\mbox{\boldmath $\theta$}$ using the available neighborhood information of sampled nodes for directed OSNs such as Pinterest, Sina microblog, and Xiami,
where a node has knowledge of in-degrees (the number of followers) and out-degrees (the number of following) of its incoming neighbors and outgoing neighbors.
For a node $v\in V$,
denote by $d^{\text{(I)}}_v$ its in-degree, $d^{\text{(O)}}_v$ its out-degree,
$\mathcal{N}^{\text{(I)}}(v)=\{u: (u,v)\in E_d\}$ the set of its followers, and $\mathcal{N}^{\text{(O)}}(v)=\{u: (v,u)\in E_d\}$ the set of its following.
Define $\psi(u, v)=0$ when $v$ is not a neighbor of $u$,
$\psi(u, v)=2$ when $v$ is an out-going and incoming neighbor of $u$, and otherwise $\psi(u, v)=1$.
Let $\mathcal{N}(v) = \mathcal{N}^{\text{(I)}}(v)\cup \mathcal{N}^{\text{(O)}}(v)$.
Using the properties of sampled nodes' neighbors, we estimate $\theta_k$ as follows
\begin{equation*}
\breve\theta_k^*=\frac{1}{\breve C_d}\sum_{i=1}^n \sum_{w\in \mathcal{N}(s_i)} \frac{\psi(s_i, w) \mathbf{1}(L(w)=l_k)}{\hat\pi_{s_i} (d^{\text{(I)}}_w+d^{\text{(O)}}_w)}, \quad 1\le k \le K,
\end{equation*}
where $\breve C_d=\sum_{i=1}^n \sum_{w\in \mathcal{N}(s_i)} \psi(s_i, w) \hat\pi_{s_i}^{-1} (d^{\text{(I)}}_w+d^{\text{(O)}}_w)^{-1}$.
\begin{theorem}\label{theorem:nodelabel2}
$\breve\theta_k^*$ ($1\le k \le K$) is an asymptotically unbiased  estimate of $\theta_k$.
\end{theorem}

\begin{pf} Applying Lemma~\ref{lemma:node}, we have
\begin{equation*}
\begin{split}
&\lim_{n\rightarrow \infty} \frac{1}{n} \sum_{i=1}^n  \left[\sum_{w\in \mathcal{N}(s_i)} \frac{\psi(s_i, w) \mathbf{1}(L(w)=l_k)}{\hat\pi_{s_i} (d^{\text{(I)}}_w+d^{\text{(O)}}_w)} \right]\\
& \xrightarrow{a.s.} \sum_{v\in V}\left(\pi_v \sum_{w\in \mathcal{N}(v)} \frac{\psi(v, w) \mathbf{1}(L(w)=l_k)}{\hat\pi_{v} (d^{\text{(I)}}_w+d^{\text{(O)}}_w)} \right)\\
&= C_\pi \sum_{v\in V}  \sum_{w\in \mathcal{N}(v)} \frac{\psi(v, w) \mathbf{1}(L(w)=l_k)}{d^{\text{(I)}}_w+d^{\text{(O)}}_w} \\
&= C_\pi \sum_{w\in V}  \sum_{v\in \mathcal{N}(w)} \frac{\psi(w, v) \mathbf{1}(L(w)=l_k)}{d^{\text{(I)}}_w+d^{\text{(O)}}_w} \\
&= C_\pi \sum_{w\in V}  \mathbf{1}(L(w)=l_k)= C_\pi |V| \theta_k.
\end{split}
\end{equation*}
The third equation holds because $\sum_{v\in \mathcal{N}(w)} \psi(w, v)= d^{\text{(I)}}_w+d^{\text{(O)}}_w$.
Similarly we proof that $\lim_{n\rightarrow \infty} \breve C^*/n \xrightarrow{a.s.} C_\pi |V| $. Therefore we have $\lim_{n\rightarrow \infty} \breve\theta_k^* \xrightarrow{a.s.} \theta_k$.
\end{pf}

Next, we propose methods for graphs such as Citeseerx website,
where we can obtain a node's neighbors' out-degrees when we sample a node.
However in-degrees of sampled nodes' neighbors are not available.
Then we estimate node label density $\theta_k$ ($1\le k \le K$) based on sampled nodes and their out-going neighbors, that is
\begin{equation*}
\breve\theta_k^{\text{(O)}}=\frac{1}{\breve C_d^*}\sum_{i=1}^n \left(\frac{\gamma \mathbf{1}(L(s_i)=l_k)}{\hat\pi_{s_i} (d^{\text{(I)}}_{s_i}+\gamma)} +\sum_{w\in \mathcal{N}^\text{(O)}(s_i)} \frac{\mathbf{1}(L(w)=l_k)}{\hat\pi_{s_i} (d^{\text{(I)}}_w+\gamma)}\right)
\end{equation*}
where $\breve C_d^*=\sum_{i=1}^n \left(\frac{\gamma}{\hat\pi_{s_i} (d^{\text{(I)}}_{s_i}+\gamma)} +\sum_{w\in \mathcal{N}^\text{(O)}(s_i)} \frac{1}{\hat\pi_{s_i} (d^{\text{(I)}}_w+\gamma)}\right)$, and $\gamma>0$.

\begin{theorem}\label{theorem:nodelabelindegree}
$\breve\theta_k^{\text{(O)}}$ ($1\le k \le K$) is an asymptotically unbiased  estimate of $\theta_k$.
\end{theorem}

\begin{pf} Denote by $V_0^{\text{(I)}}$ the set of nodes in $V$ whose in-degrees are larger than 0.
Clearly only nodes in $V_0^{\text{(I)}}$ can appear in a node's out-going neighbor list.
Applying Lemma~\ref{lemma:node}, we have
\begin{equation*}
\begin{split}
&\lim_{n\rightarrow \infty} \frac{1}{n} \sum_{i=1}^n  \left[\frac{\gamma \mathbf{1}(L(s_i)=l_k)}{\hat\pi_{s_i} (d^{\text{(I)}}_{s_i}+\gamma)} +\sum_{w\in \mathcal{N}^\text{(O)}(s_i)} \frac{\mathbf{1}(L(w)=l_k)}{\hat\pi_{s_i} (d^{\text{(I)}}_w+\gamma)} \right]\\
& \xrightarrow{a.s.} \sum_{v\in V}\pi_v \left(\frac{\gamma \mathbf{1}(L(v)=l_k)}{\hat\pi_v (d^{\text{(I)}}_v+\gamma)} +\sum_{w\in \mathcal{N}^\text{(O)}(v)} \frac{\mathbf{1}(L(w)=l_k)}{\hat\pi_v (d^{\text{(I)}}_w+\gamma)} \right)\\
&= C_\pi \left( \sum_{v\in V}  \frac{\gamma \mathbf{1}(L(v)=l_k)}{d^{\text{(I)}}_v+\gamma} + \sum_{v\in V\backslash V_0^{\text{(I)}}} \frac{d^{\text{(I)}}_v \mathbf{1}(L(v)=l_k)}{d^{\text{(I)}}_v+\gamma} \right) \\
&= C_\pi \left( \sum_{v\in V_0^{\text{(I)}}}  \mathbf{1}(L(v)=l_k) + \sum_{v\in V\backslash V_0^{\text{(I)}}} \mathbf{1}(L(v)=l_k) \right) \\
&= C_\pi \sum_{w\in V}  \mathbf{1}(L(w)=l_k)= C_\pi |V| \theta_k.
\end{split}
\end{equation*}
The first equation holds because
\begin{equation*}
\begin{split}
&\sum_{v\in V}\sum_{w\in \mathcal{N}^\text{(O)}(v)} \frac{\mathbf{1}(L(w)=l_k)}{d^{\text{(I)}}_w+\gamma}\\
&=\sum_{w\in V\backslash V_0^{\text{(I)}}} \sum_{v\in \mathcal{N}^\text{(O)}(w)} \frac{\mathbf{1}(L(w)=l_k)}{d^{\text{(I)}}_w+\gamma}\\
&=\sum_{w\in V\backslash V_0^{\text{(I)}}} \frac{d^{\text{(I)}}_w \mathbf{1}(L(w)=l_k)}{d^{\text{(I)}}_w+\gamma}.
\end{split}
\end{equation*}
Similarly we proof that $\lim_{n\rightarrow \infty} X/n \xrightarrow{a.s.} C_\pi |V| $. Therefore we have $\lim_{n\rightarrow \infty} \breve\theta_k^{\text{(O)}} \xrightarrow{a.s.} \theta_k$.
\end{pf}

Similar to the mixture estimator~(\ref{eq:mixestimator}), $\breve\theta_k^*$ and $\breve\theta_k^{\text{(O)}}$ can be combined with $\hat\theta_k$ to estimate $\theta_k$ more accurately.

%

\section{Edge Label Density Estimation} \label{sec:edgelabeldensity}
Define $L(u,v)$ to be the label of edge $(u,v)$, with range $\mbox{\boldmath $L'$}=\{l_1',...,l_{K'}'\}$.
Denote the edge label density by $\btau=(\tau_1,\ldots,\tau_{K'})$, where $\tau_k$ ($1\le k \le K'$) is the fraction of edges with label $l_k'$.
For undirected graph $G$, we let edge label function $L(u,v)=L(v,u)$.
For example, when define $L(u,v)=(\min\{d_u, d_v\},\max\{d_u, d_v\})$ for edge $(u,v)$ in undirected graph $G$, the pair of degrees of nodes $u$ and $v$, $\btau$ is the joint node degree distribution. Note that the labels of edges $(u,v)$ and $(v,u)$ in directed graph $G_d$ may not be the same.
In this section we propose methods for estimating $\btau$ for undirected graphs and directed graphs respectively.

\subsection{Simple Estimators of Edges Densities}
Based on edges $\{(u_i, v_i)\}_{i=1,\ldots,n}$ sampled by RW and FS,
~\cite{IMC_Ribeiro10} estimates $\tau_k$ for an undirected graph $G$ as follows
\begin{equation}\label{eq:firstestimatortau}
\hat\tau_k=\frac{1}{n}\sum_{i=1}^n \mathbf{1}(L'(u_i, v_i)=l_k'), \quad 1\le k \le K'.
\end{equation}
It shows that $\hat\tau_k$ ($1\le k \le K'$) is an asymptotically unbiased estimate of $\tau_k$ for undirected graphs.
Similarly, in this paper we estimate $\btau$ of directed graph $G_d$ as follows
\begin{equation*}
\begin{split}
\hat\tau_k^*=\frac{1}{H_d}\sum_{i=1}^n &\left(\mathbf{1}(L'(u_i, v_i)=l_k') \mathbf{1}((u_i, v_i)\in E_d)\right.\\
&\left. +\mathbf{1}(L'(v_i, u_i)=l_k') \mathbf{1}((v_i, u_i)\in E_d)\right).
\end{split}
\end{equation*}
where $H_d=\sum_{i=1}^n \mathbf{1}((u_i, v_i)\in E_d) + \mathbf{1}((v_i, u_i)\in E_d)$.
We can easily prove $\hat\tau_k^*$ ($1\le k \le K'$) is an asymptotically unbiased estimate of $\tau_k$ for directed graph $G_d$.

\subsection{Estimators Using Neighborhood Information of Sampled Nodes}
In this paper we assume that we can obtain the labels of all (incoming and outgoing) edges of a node when we query a node from $G$ ($G_d$).
Using the neighborhood information of sampled nodes $s_i$ ($1\le i\le n$) obtained by UNI, RW and FS, we estimate $\tau_k$ of $G$ as follows
\begin{equation}\label{eq:secondestimatortau}
\breve\tau_k=\frac{1}{\breve H}\sum_{i=1}^n \sum_{w\in \mathcal{N}(s_i)} \frac{\mathbf{1}(L'(s_i, w)=l_k')}{\hat\pi_{s_i}}, \quad 1\le k \le K',
\end{equation}
where $\breve H =\sum_{i=1}^n \sum_{w\in \mathcal{N}(s_i)} {\hat\pi_{s_i}}^{-1}$. Then we have

\begin{theorem}\label{theorem:edgelabel}
$\breve\tau_k$ ($1\le k \le K'$) is an asymptotically unbiased estimate of $\tau_k$ for undirected graphs.
\end{theorem}
\begin{pf} Applying Lemma~\ref{lemma:node}, we have
\begin{equation*}
\begin{split}
&\lim_{n\rightarrow \infty} \frac{1}{n} \sum_{i=1}^n \sum_{w\in \mathcal{N}(s_i)} \frac{\mathbf{1}(L'(s_i, w)=l_k')}{\hat\pi_{s_i}}\\
&\xrightarrow{a.s.} \sum_{v\in V} \left(\pi_v \sum_{w\in \mathcal{N}(v)} \frac{\mathbf{1}(L'(v, w)=l_k')}{\hat\pi_v} \right)\\
&=C_\pi \sum_{v\in V} \sum_{w\in \mathcal{N}(v)} \mathbf{1}(L'(v, w)=l_k')= 2C_\pi |E|\tau_k.
\end{split}
\end{equation*}
Similarly we have $\lim_{i\rightarrow \infty} \text{E}[\breve H/n]\rightarrow 2 C_\pi |E|$.
Therefore we have $\lim_{n\rightarrow \infty} \breve\tau_k \xrightarrow{a.s.} \tau_k$.
\end{pf}

Utilizing the free neighborhood information of sampled nodes $s_i$ ($1\le i\le n$), we estimate $\tau_k$ of $G_d$ as follows
\begin{equation*}
\begin{split}
\breve\tau_k^*=\frac{1}{\breve H_d}\sum_{i=1}^n \sum_{w\in \mathcal{N}(s_i)} &\left( \frac{\mathbf{1}(L'(s_i, w)=l_k') \mathbf{1}((s_i, w)\in E_d)}{\hat\pi_{s_i}}\right.\\
&\left. +\frac{\mathbf{1}(L'(w, s_i)=l_k') \mathbf{1}((w, s_i)\in E_d)}{\hat\pi_{s_i}}\right)
\end{split}
\end{equation*}
where $\breve H_d =\sum_{i=1}^n \sum_{w\in \mathcal{N}(s_i)} \frac{\mathbf{1}((s_i, w)\in E_d)+\mathbf{1}((w, s_i)\in E_d)}{\hat\pi_{s_i}}$.
Similar to Theorem~\ref{theorem:edgelabel}, we have
\begin{theorem}\label{theorem:directededgelabel2}
$\breve\tau_k^*$ ($1\le k \le K'$) is an asymptotically unbiased  estimate of $\tau_k$ for directed graphs.
\end{theorem}

In summary, $\hat\btau=(\hat\tau_1,\ldots,\hat\tau_{K'})$ and $\hat\btau^*=(\hat\tau_1^*,\ldots,\hat\tau_{K'}^*)$ computed as described above form asymptotically unbiased estimates of $\btau$  for undirected and directed graphs respectively.
When properties of sampled nodes' neighbors are available,
we utilize all edge labels observed from this neighborhood information,
and provide asymptotically unbiased estimates $\breve\btau=(\breve\tau_1,\ldots,\breve\tau_{K'})$ and $\breve\btau^*=(\breve\tau_1^*,\ldots,\breve\tau_{K'}^*)$ of $\btau$ for undirected and directed graphes respectively.

\section{High Degree Node Detection} \label{sec:supernode}
In this section, we study the problem of detecting the $N$ nodes with the largest degrees in undirected graph $G=(V,E)$.
Let $S$ be the set of nodes sampled by methods such as RW.
Previous methods use the $N$ nodes in $S$ with the largest degrees to estimate high degree nodes~\cite{Lim2011,Cooper2012}.
In~\cite{Cooper2012}, weighted RW (WRW) is used to detect high degree nodes.
WRW can be viewed as a RW over a weighted graph, where each edge $(u,v)\in E$ has a positive weight $w(u,v)=w(v,u)$~\cite{Coppersmith1993}. At each step, WRW selects the next-hop node $v$ at random among the neighbors of the
current node $u$ with probability proportional to weight $w(u,v)$.
WRW (with well defined edge weights) and RW are efficient for detecting high degree nodes, since they are biased to sample high degree nodes~\cite{IMC_Ribeiro10,Cooper2012}.
Note that the WRW proposed in~\cite{Cooper2012} sets weight $w(u,v)=(d_u d_v)^\beta$ for each edge $(u,v)\in E$ for detecting top-$N$ high degree nodes, which indicates that at each step their WRW need to obtain degrees of current sampled node's neighbors.
However their description does not account for the cost of retrieving this information.
In~\cite{Maiya2010}, a new method, expansion sampling (XS), is proposed for detecting high degree nodes.
Denote by $\mathcal{N}(S)$ the neighborhood of $S$,
where $\mathcal{N}(S)$ consists of nodes in $V\setminus S$ that are neighbors of nodes in $S$, that is $\mathcal{N}(S)=\{u: u\in V\setminus S, \exists v\in S, (u,v)\in E \}$.
Starting from a random node $s$, and $S=\{s\}$, XS adds the node in $\mathcal{N}(S)$ which has the largest number of neighbors in $V\setminus (\mathcal{N}(S)\cup S)$ to $S$, and repeats this process.
For a node $u\in \mathcal{N}(S)$, denote by $d_u^{(S)}$ the number of edges between $u$ and nodes in $S$, and $d_u^{(\mathcal{N}(S))}$  the number of edges between $u$ and nodes in $\mathcal{N}(S)$.
Then, the number of its neighbors in $V\setminus (\mathcal{N}(S)\cup S)$ equals $d_u-d_u^{S}-d_u^{(\mathcal{N}(S))}$.
From knowledge of edges of nodes in $S$, we know $d_u$ and $d_u^{S}$.
However $d_u^{(\mathcal{N}(S))}$ cannot be obtained based on available information of $S$ and $\mathcal{N}(S)$.
In order to identify the node in $\mathcal{N}(S)$ which has the largest number of neighbors in $V\setminus (\mathcal{N}(S)\cup S)$, it is necessary to crawl all nodes in $\mathcal{N}(S)$.
The original description of XS~\cite{Maiya2010} does not account for this cost.
On the other hand when each node has knowledge of its neighbors' degrees.
It is possible to identify the node in $\mathcal{N}(S)$ that has the largest number of neighbors in $V\setminus S$.
Therefore we propose a Modified XS (MXS) method, which adds this node to $S$ at each step.
Finally we output the $N$ nodes with the largest degrees in $\mathcal{N}(S)\cup S$ as the final results.
The above methods can be easily modified to identify $N$ nodes with the largest in-degrees or out-degrees in directed graph such as Sina microblog, Tencent microblog, and Xiami, where a node has knowledge of its neighbors' out-degrees and in-degrees. Here at each step MXS adds the node with the largest sum of in-degree and out-degree in $\mathcal{N}(S)$ to $S$.

\section{Short Path Discovery} \label{sec:shortpath}
In this section, we study the problem of performing topology discovery and message routing
with incomplete topological information, which is important for applications such as discovery of short paths between OSN users
and routing algorithms (e.g. Bubble Rap~\cite{Hui2008}) for delivering messages between users using a social network.
Formally the problem is: Two nodes $u$ and $v$ are looking for short paths on undirected graph $G$.
Ribeiro et al.~\cite{RibeiroNetSci2012} find that a RW has the ability to observe a large fraction of the edges by visiting a relatively small number of nodes on power law graphs. Here an edge is observed when at least one of its endpoints is visited by the RW. They propose a RW based short path discovery algorithm works as follows: Two RWs are started from $u$ and $v$ separately. Each RW takes $B$ steps.
Let $S$ be the set of nodes sampled by two RWs. Finally They use the shortest path in observed graph $G^*=(V^*, E^*)$ for routing between $u$ and $v$, where $V^*=S\cup \mathcal{N}(S)$ and $E^*$ consists of edges in $E$ which have at least one endpoint contained by $S$.
From Section~\ref{sec:supernode}, we know that WRW and MXS can
efficiently find high degree nodes and observe more edges based on neighborhood information.
We propose two new methods, which perform a WRW and MXS starting from two initial nodes $u$ and $v$ respectively.
We finally use the shortest path in graph $G^*=(V^*, E^*)$ observed by WRW or MXS for routing between $u$ and $v$.

\section{Data Evaluation} \label{sec:results}
We perform our experiments on a variety of real world networks
that are summarized in Table~\ref{tab:datasets}.
Xiami is a popular website
devoted to music streaming and music recommendations.
Similar to Twitter, Xiami
builds a social network based on follower and following relationships.
Flickr and YouTube are popular photo sharing and video sharing websites.
In these websites, a user can subscribe to other user updates such as blogs and photos.
These networks can be represented by direct graphs, with
nodes representing users and a directed edge from $u$ to $v$
represents that user $u$ subscribes to user $v$.
Epinions is a who-trusts-whom OSN providing general consumer reviews, where a directed edge from $u$ to $v$ represents that user $u$ trusts user $v$.
Slashdot is a technology-related news website for its specific user community, where a directed edge from $u$ to $v$ represents that user $u$ tags user $v$ as a friend or foe.
In the following experiments, we evaluate our methods in comparisons with previous methods based on the largest connected component (LCC) of these graphs under the same sampling budget $B$, where $B$ is defined as the number of sampled nodes.

\begin{table}[htb]
\begin{center}
\caption{Overview of graph datasets used in our simulations.``directed-edges" refers to the number of directed edges in a
directed graph, ``edges" refers to the number of edges in an
undirected graph, and ``LCC" refers to the largest connected component
of a given graph.\label{tab:datasets}}
\begin{tabular}{|c|ccc|}
\hline
\multirow{2}{*}{Graph} &  & LCC & \\
\cline{2-4}
&nodes&edges&directed-edges\\
\hline
Xiami~\cite{PinghuiContent2012}&1,748,010&16,015,779&16,568,449\\
YouTube~\cite{MisloveIMC2007}&1,134,890&2,987,624&4,942,035\\
Flickr~\cite{MisloveIMC2007}&1,624,992&15,476,835&22,477,014\\
Soc-Epinions~\cite{Richardson2003}&75,877&405,739&405,739\\
Soc-Slashdot~\cite{LeskovecIM2009}&77,360&469,180&828,161\\
\hline
\end{tabular}
\end{center}
\end{table}

\subsection{Node Label Density Estimation}
Let $\mbox{\boldmath $\theta$}=(\theta_1,\ldots,\theta_{K})$ be the (in-) degree distribution,
where $\theta_k$ ($1\le k \le K$) is the fraction of nodes with (in-) degree $k$.
In our study, we estimate
both $\theta_k$ and $\xi_k=\sum_{i=k+1}^K \theta_i$, the CCDF (complementary cumulative
distribution function) of $\mbox{\boldmath $\theta$}$, which is the statistic of choice when it comes to display (in-degree) degree distributions.
For estimator $\hat \theta_k$, we define the normalized root mean square error (NMSE) as
$\text{NMSE}(\hat{\theta}_k)=\sqrt{\text{E}[(\hat{\theta}_k-\theta_k)^2]}/\theta_k$, $k=1,2,\dots$.
In the following experiments,
we use 1,000 independent runs to estimate $\text{E}[(\hat{\theta}_k-\theta_k)^2]$.
Similarly, we define the NMSE of the CCDF of $\mbox{\boldmath $\theta$}$, which we denote
as the CNMSE to avoid confusion with the NMSE of $\mbox{\boldmath $\theta$}$.

\begin{figure}[htb]
\center
\subfigure[Xiami]{
\includegraphics[width=0.23\textwidth]{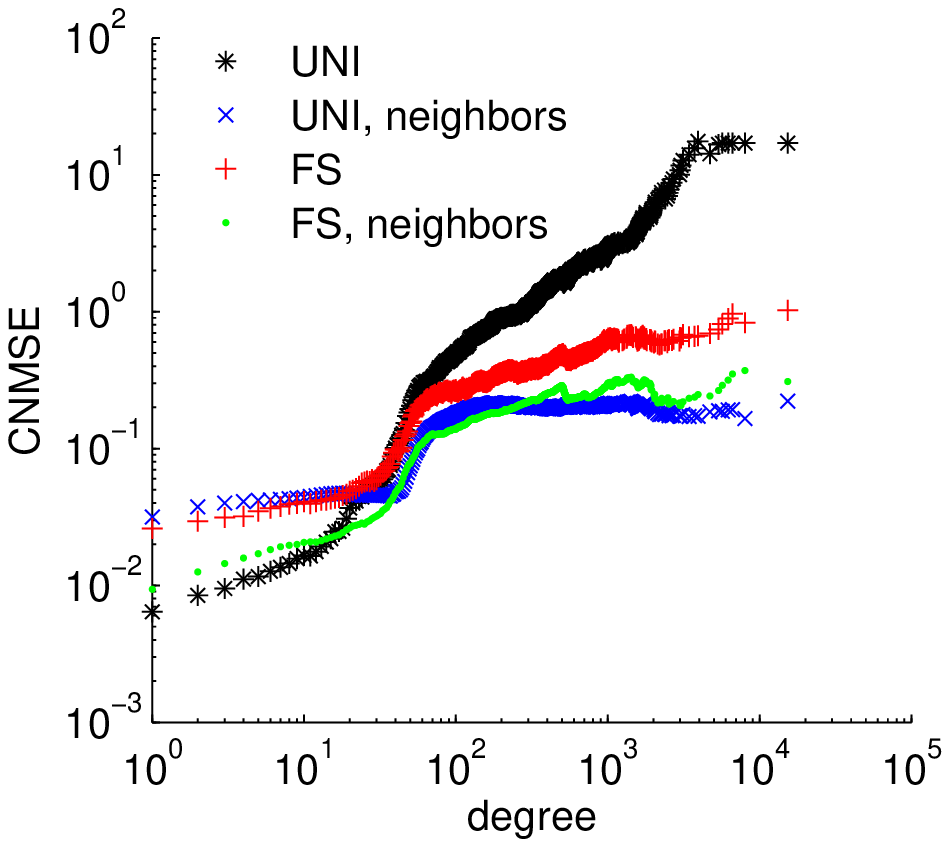}}
\subfigure[YouTube]{
\includegraphics[width=0.23\textwidth]{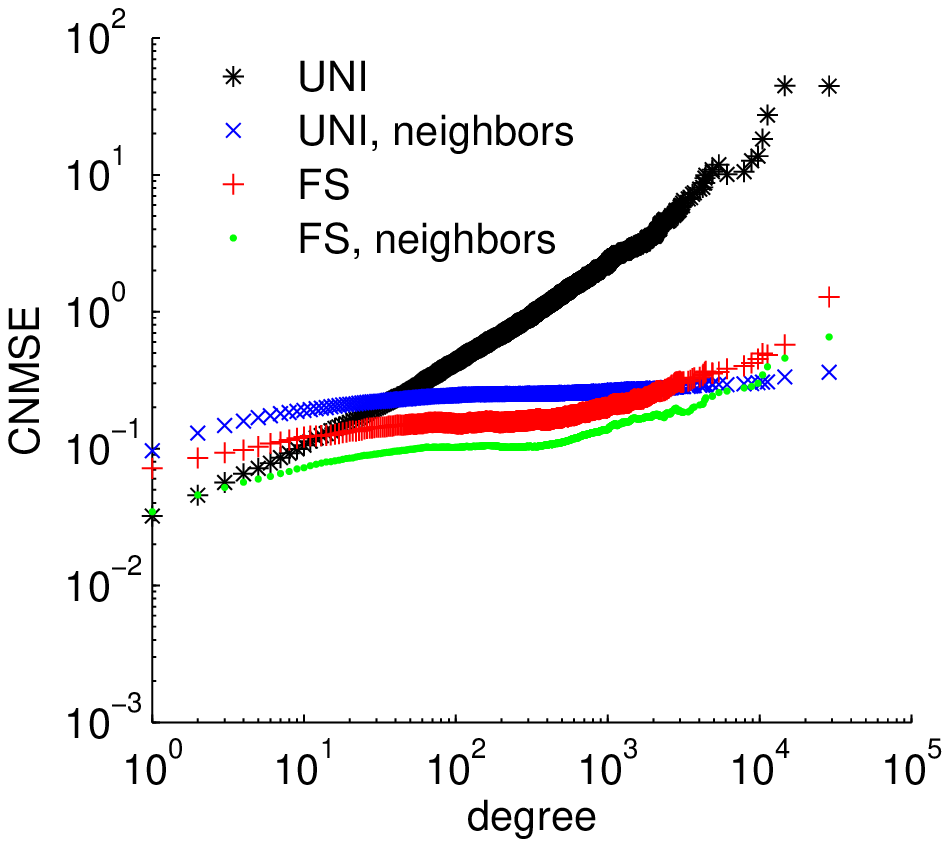}}
\caption{Results of degree distribution estimations for undirected graphs, $B=0.001|V|$.}\label{fig:nodelabeldensityunirwcmp}
\end{figure}

\begin{figure}[htb]
\center
\subfigure[Xiami]{
\includegraphics[width=0.23\textwidth]{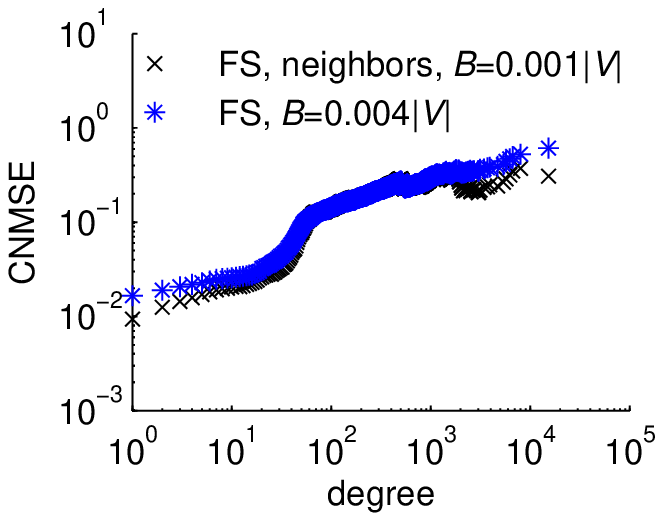}}
\subfigure[YouTube]{
\includegraphics[width=0.23\textwidth]{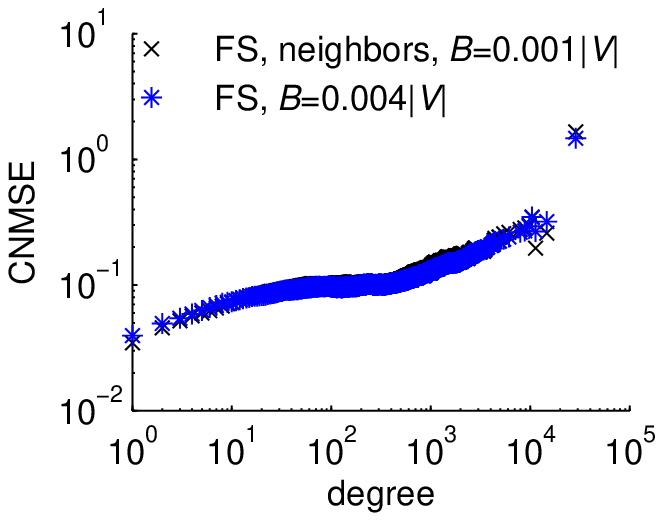}}
\caption{To achieve the same MSE, regular FS requires at least $4\times$ the number of the samples of FS  with neighbor information.}\label{fig:nodelabeldensityfssameMSE}
\end{figure}

Fig.~\ref{fig:nodelabeldensityunirwcmp} shows the CNMSEs of estimates of degree distribution $\mbox{\boldmath $\theta$}=(\theta_1,\ldots,\theta_{K})$, where sampling budget $B=0.001|V|$.
Fig.~\ref{fig:nodelabeldensityunirwcmp} shows that the degree estimates produced by UNI and FS using neighbor information almost have the same accuracy. For FS, the degree distribution estimate greatly improves when neighbor information is used, which is almost {\em twice} as accurate than previous FS without using neighbor information for Xiami.
~\cite{IMC_Ribeiro10,INFOCOM_Ribeiro2012} show that NMSEs are roughly proportional to $1/\sqrt{B}$.
It indicates that FS using neighbor information is {\em four times} more time efficient than the previous FS,
which is consistent with our results shown in Fig.~\ref{fig:nodelabeldensityfssameMSE}.
For UNI method, the degree distribution estimator based on using the neighbor information of sampled nodes exhibits larger errors than the estimator given by sampled nodes for small degrees (degrees smaller than 20 for Xiami, 30 for YouTube).
For the degree distribution estimator given by neighbors of sampled nodes, we can see that FS is more accurate than UNI for most degrees.

\begin{figure}[htb]
\center
\subfigure[Xiami, UNI]{
\includegraphics[width=0.23\textwidth]{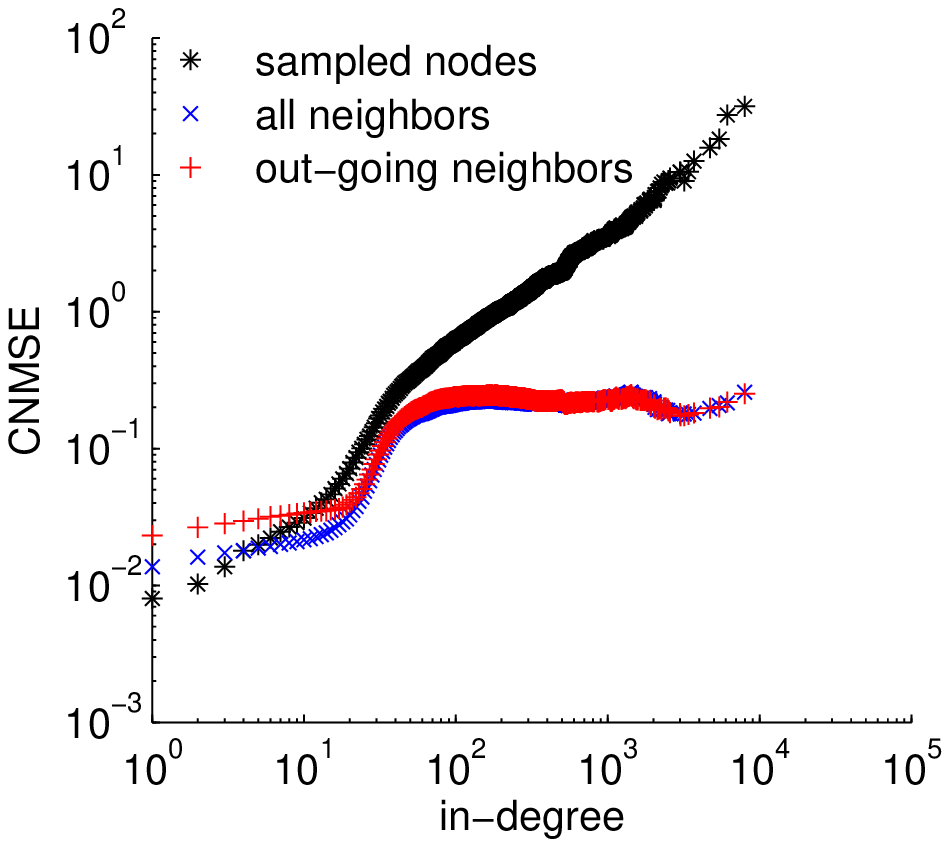}}
\subfigure[Xiami, FS]{
\includegraphics[width=0.23\textwidth]{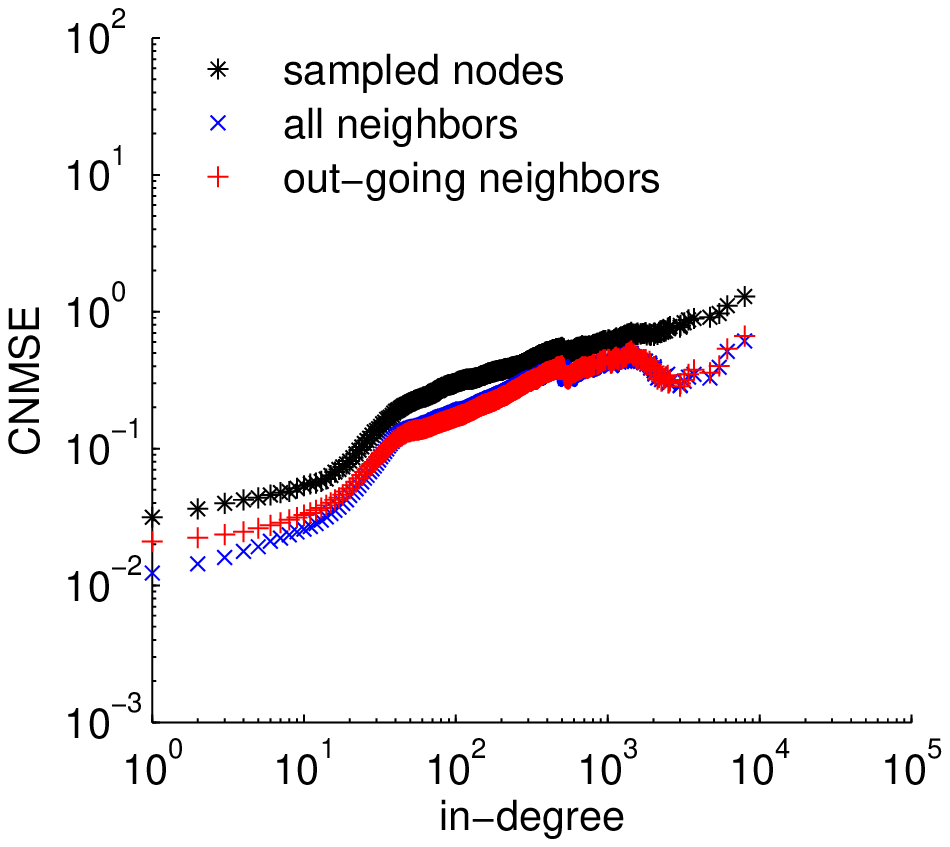}}
\subfigure[YouTube, UNI]{
\includegraphics[width=0.23\textwidth]{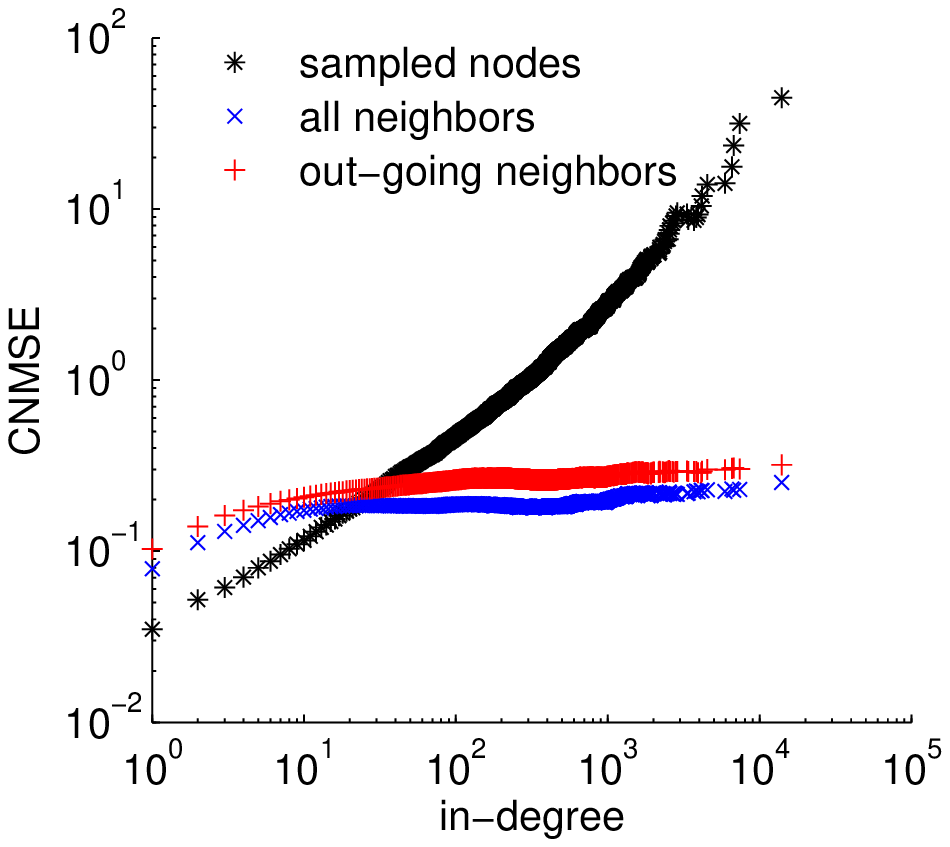}}
\subfigure[YouTube, FS]{
\includegraphics[width=0.23\textwidth]{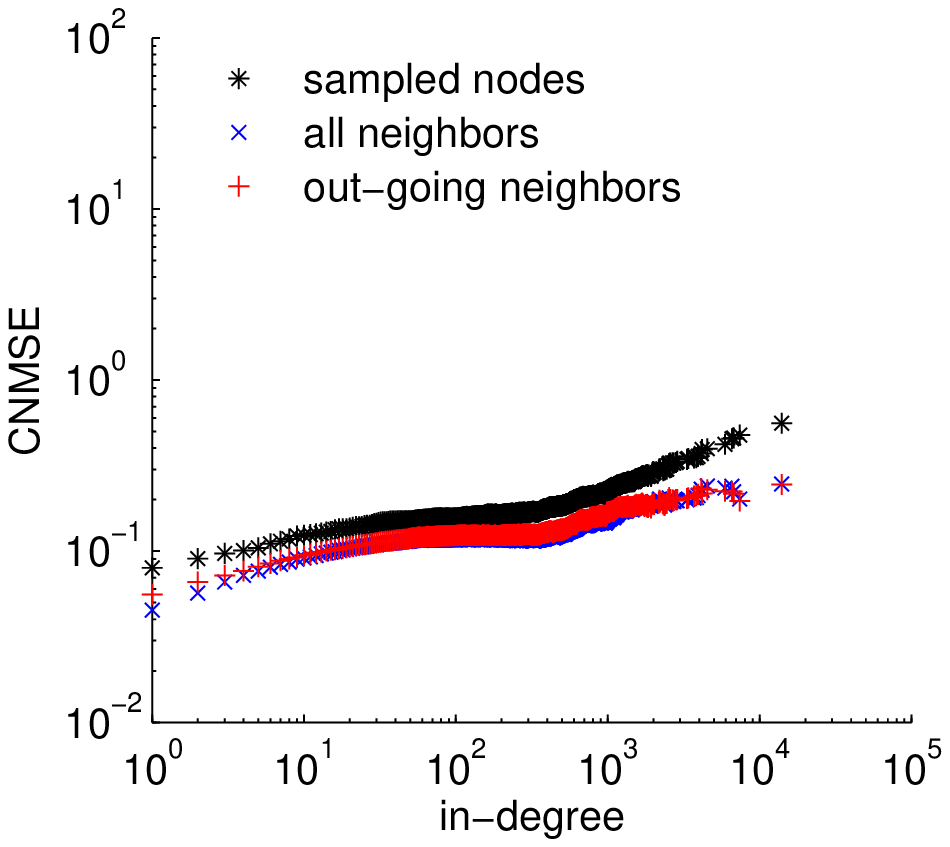}}
\caption{Results of in-degree distribution estimations for directed graphs, $B=0.001|V|$.}\label{fig:dirnodelabeldensityunirwcmp}
\end{figure}

\begin{figure}[htb]
\center
\subfigure[Xiami, UNI]{
\includegraphics[width=0.23\textwidth]{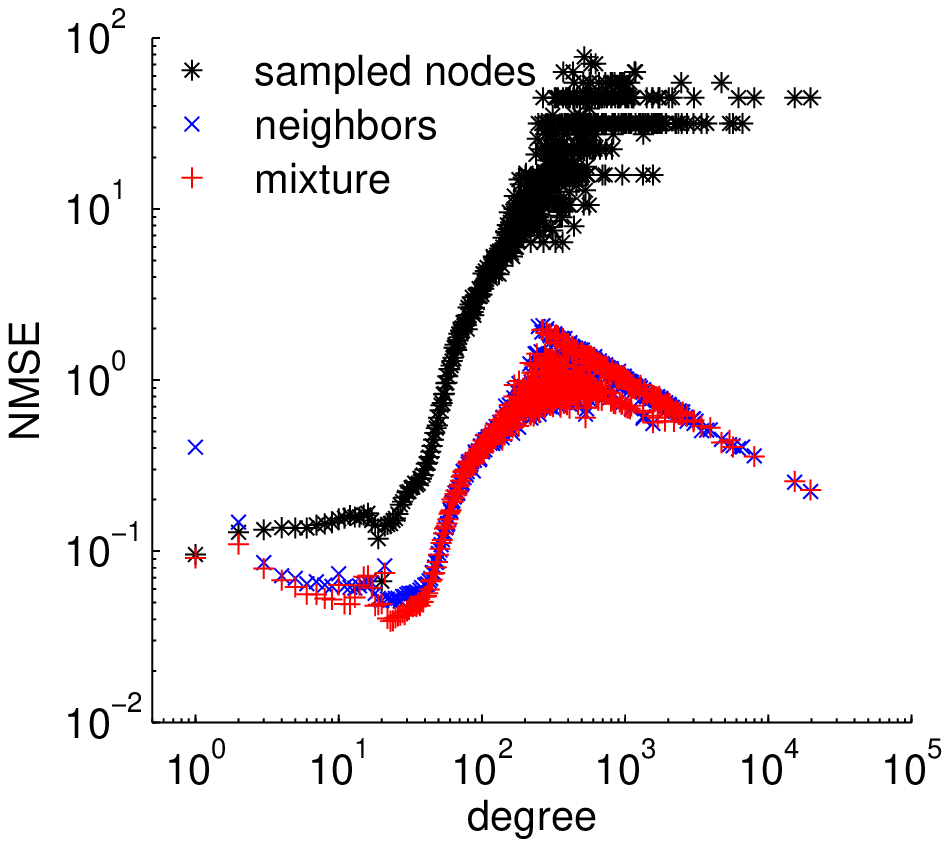}}
\subfigure[Xiami, RW]{
\includegraphics[width=0.23\textwidth]{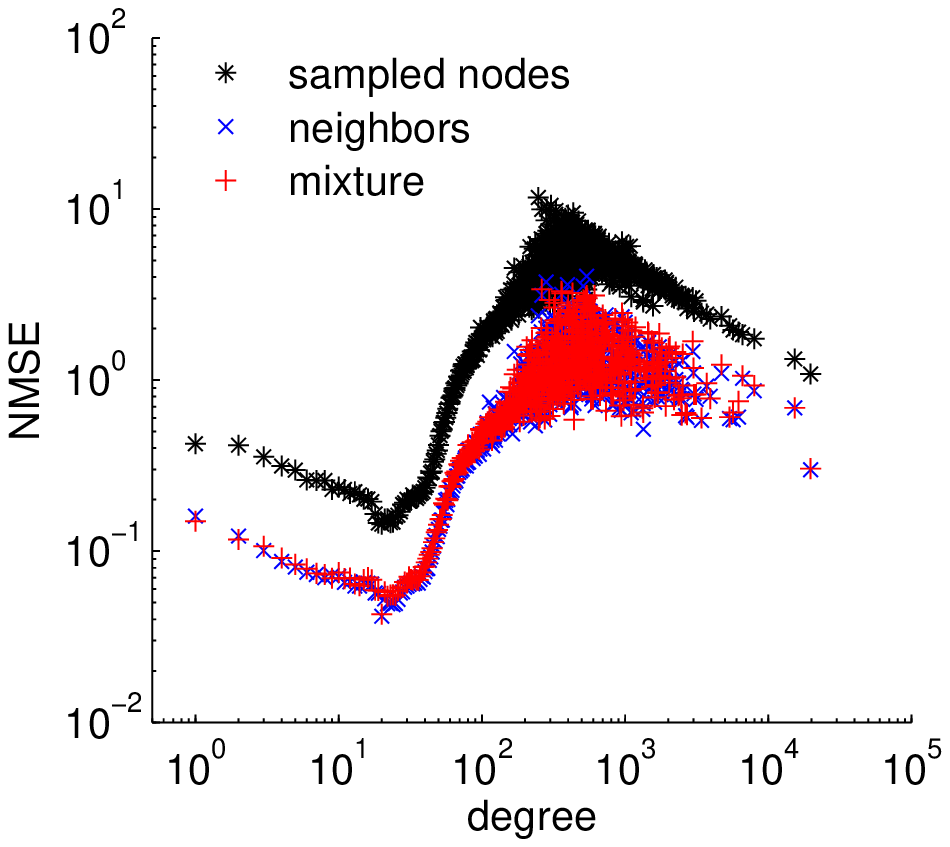}}
\subfigure[YouTube, UNI]{
\includegraphics[width=0.23\textwidth]{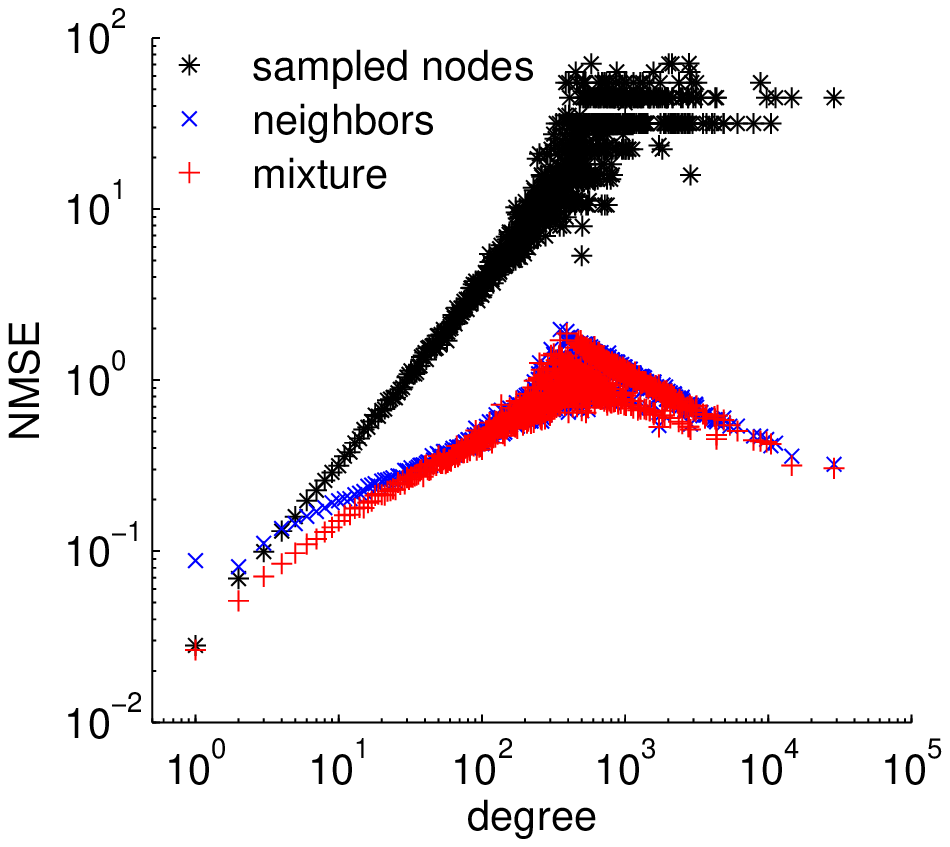}}
\subfigure[YouTube, RW]{
\includegraphics[width=0.23\textwidth]{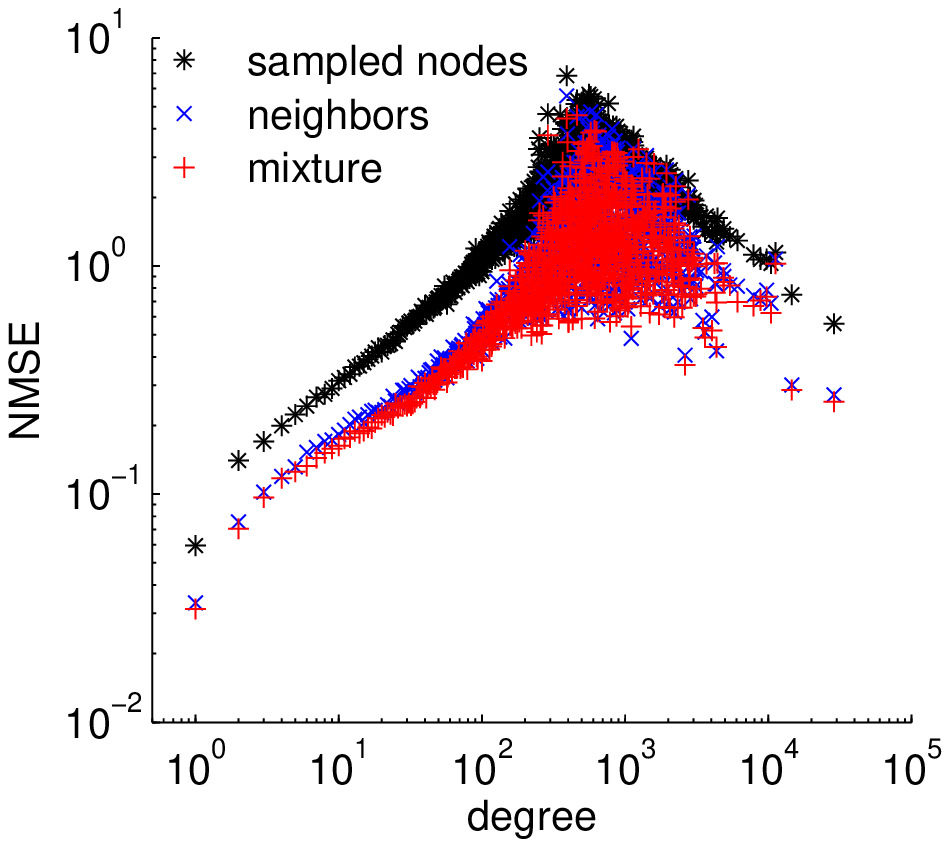}}
\caption{Results of degree distribution estimations for the mixture estimator, $B=0.001|V|$.}\label{fig:nodelabeldensitymix}
\end{figure}

\begin{figure}[htb]
\center
\subfigure[Flickr]{
\includegraphics[width=0.23\textwidth]{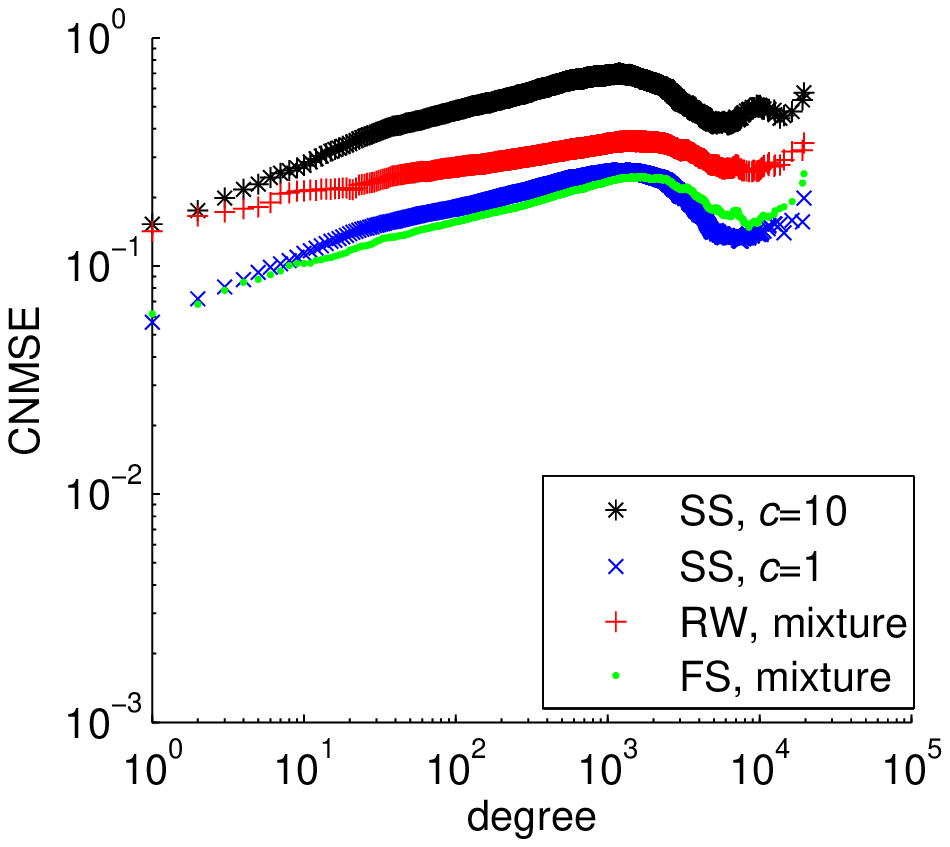}}
\subfigure[Xiami]{
\includegraphics[width=0.23\textwidth]{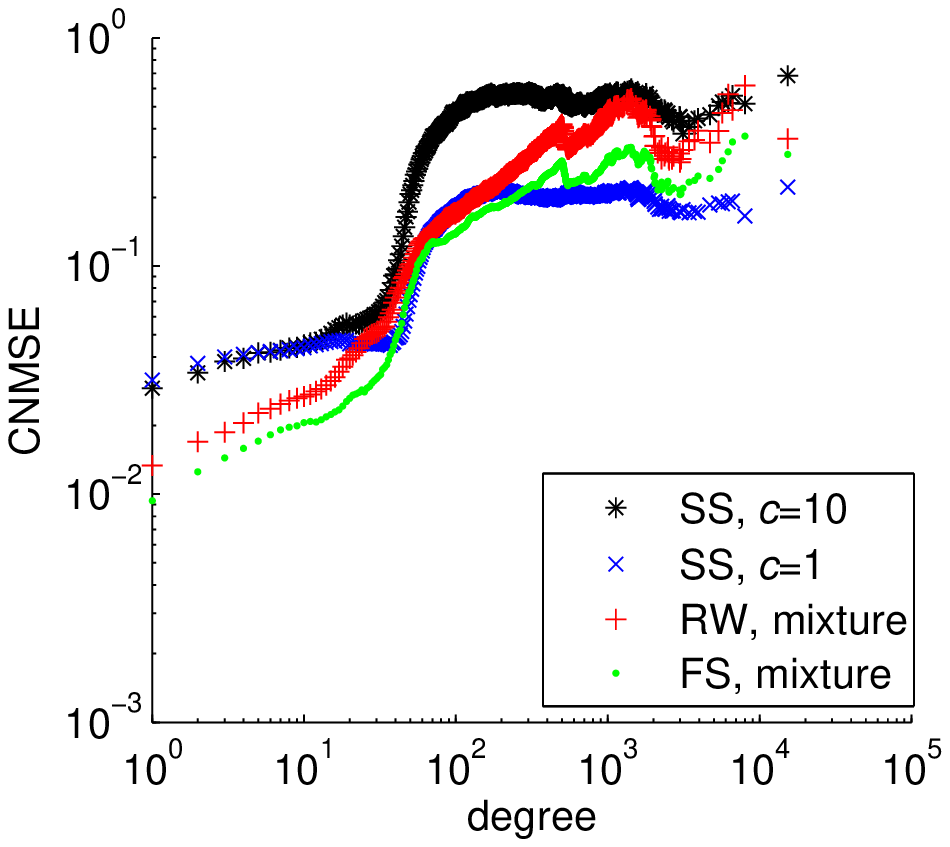}}
\caption{Results of degree distribution estimations for different node sampling cost $c$, $B=0.001|V|$.}\label{fig:cmpkkd12}
\end{figure}

For directed graphs, Fig.~\ref{fig:dirnodelabeldensityunirwcmp} shows results for the in-degree distribution estimates.
When in-degrees and out-degrees of sampled nodes are available, the in-degree distribution estimator given by neighbors of sampled nodes {\em outperforms} the estimator given by sampled nodes for FS method.
For small in-degrees (3 for Xiami, 18 for YouTube), the in-degree distribution estimator given by neighbors of sampled nodes exhibits larger errors than the estimator given by sampled nodes for UNI method.
Meanwhile, the results show that we can also give an accurate in-degree distribution estimate given by out-going neighbors of sampled nodes, which is a little less accurate than the estimate obtained by all neighbors' information.
Fig.~\ref{fig:nodelabeldensitymix} shows the results of the mixture estimator
in~(\ref{eq:mixestimator}).
We observe that the mixture estimator {\em outperforms} the estimator based on sampled nodes and the estimator based on neighbors of sampled nodes.
Let $c$ denote the cost of UNI, the average number of IDs queried until one valid ID is obtained. For example, Flickr has a random node sampling
cost of $c = 77$~\cite{INFOCOM_Ribeiro2012}. Here we set the cost of crawling methods FS and RW as 1.
Next we compare with performance of crawling methods with social sampling (SS), a node sampling method proposed by Dasgupta et al.~\cite{Dasgupta2012}.
Here SS is equivalent to the estimator given by neighbors of nodes sampled by UNI.
Fig.~\ref{fig:cmpkkd12} shows that SS exhibits larger errors as $c$ increases.
When sampling cost $c=10$, FS and RW are much more accurate than SS under the same sampling budget. Meanwhile we can see that FS exhibits smaller errors than RW.

\subsection{Edge Label Density Estimation}
We evaluate the performance of our methods for estimating
the joint degree distribution $\bphi=(\phi(i,j): i\ge j> 0)$ for undirected graph $G$, where $\phi(i,j)$ is the fraction of edges consisting of two nodes with degree $i$ and $j$ separately.
For two-dimensional distribution $\bphi$, we define $\delta$ as
\[ \delta=\sqrt{\sum_{i\ge j> 0} (\hat\phi(i,j)-\phi(i,j))^2} \,\, , \]
which is a metric that measures the error of its estimate $\hat\bphi$.

\begin{figure}[htb]
\center
\subfigure[Soc-Epinions]{
\includegraphics[width=0.23\textwidth]{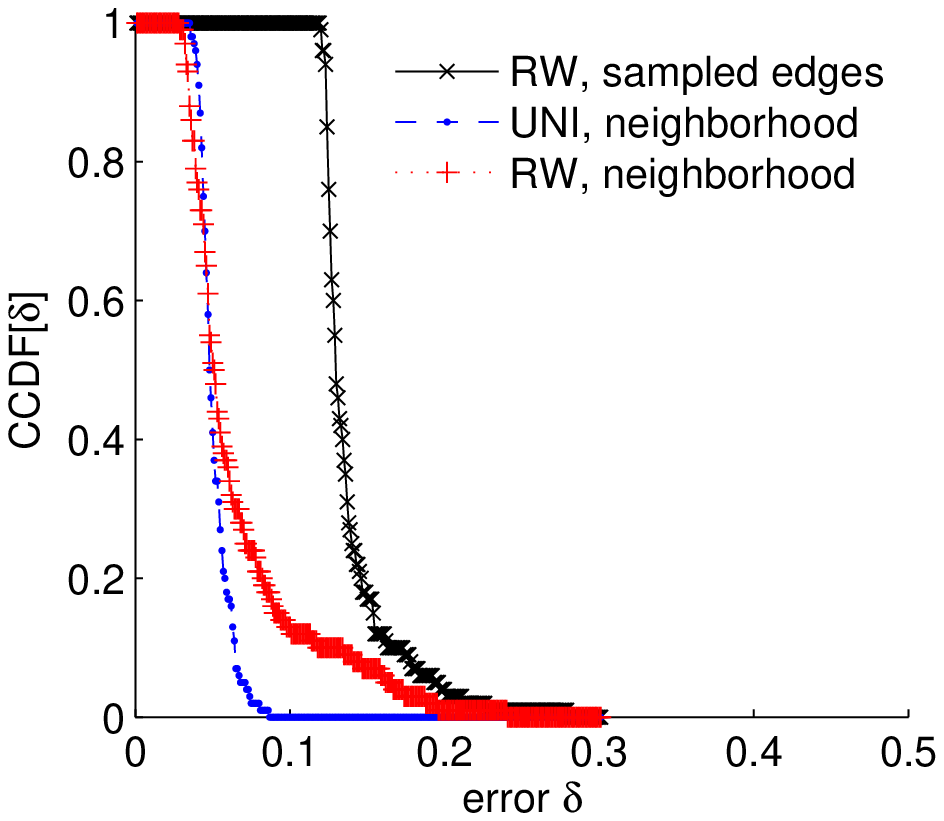}}
\subfigure[Soc-Slashdot]{
\includegraphics[width=0.23\textwidth]{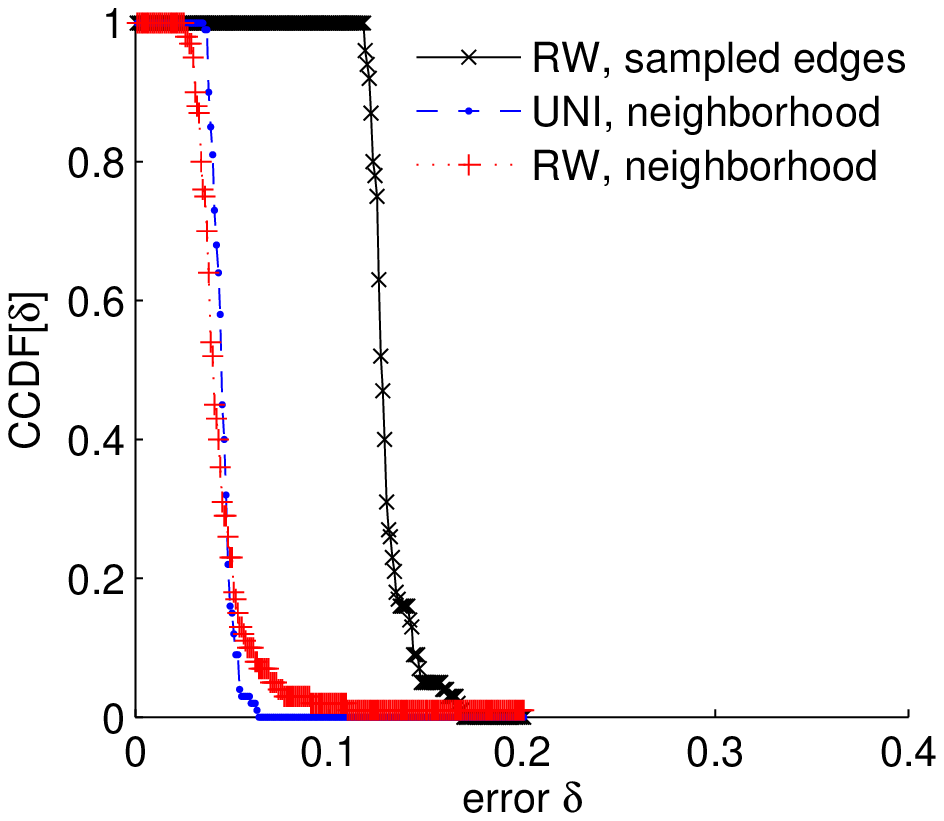}}
\caption{CCDF of errors of joint degree distribution estimates, $B=0.001|V|$.}\label{fig:jddcmp}
\end{figure}

Fig.~\ref{fig:jddcmp} shows the complementary cumulative distribution function (CCDF) of $\delta$ for 1,000 independent estimates, where the sampling budget is $B=0.001|V|$.
It shows that RW and UNI using sampled nodes' neighborhood information are more accurate.
All estimates have errors larger than 0.1 when we have no knowledge of sampled nodes' degrees.
More than 85\% of estimates have errors smaller than 0.1 when sampled nodes' degrees are available.

\begin{figure}[htb]
\begin{center}
\includegraphics[width=0.4\textwidth]{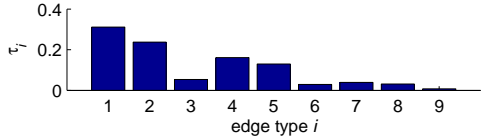}
\caption{(Xiami) Density of edges with different types. Type 1: M$\rightarrow$M, 2: M$\rightarrow$F, 3: M$\rightarrow$U,
4: F$\rightarrow$M, 5: F$\rightarrow$F, 6: F$\rightarrow$U,
7: U$\rightarrow$M, 8: U$\rightarrow$F, 9: U$\rightarrow$U.}\label{fig:genderedgedensity}
\end{center}
\end{figure}

\begin{figure}[htb]
\center
\subfigure[$B=0.001|V|$]{
\includegraphics[width=0.23\textwidth]{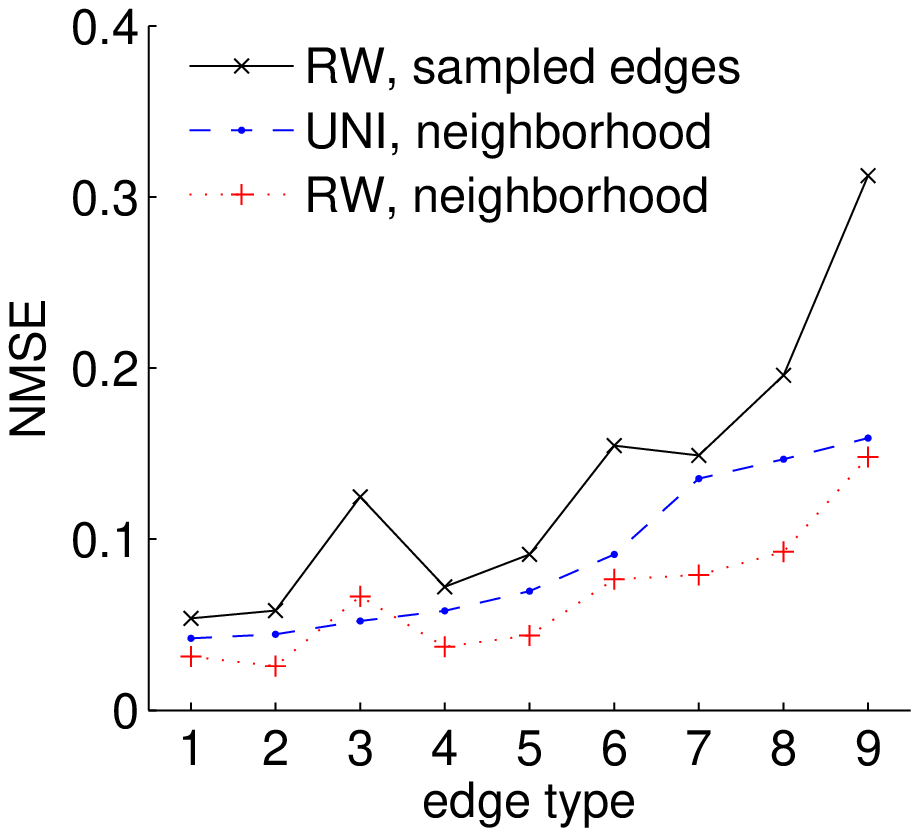}}
\subfigure[$B=0.01|V|$]{
\includegraphics[width=0.23\textwidth]{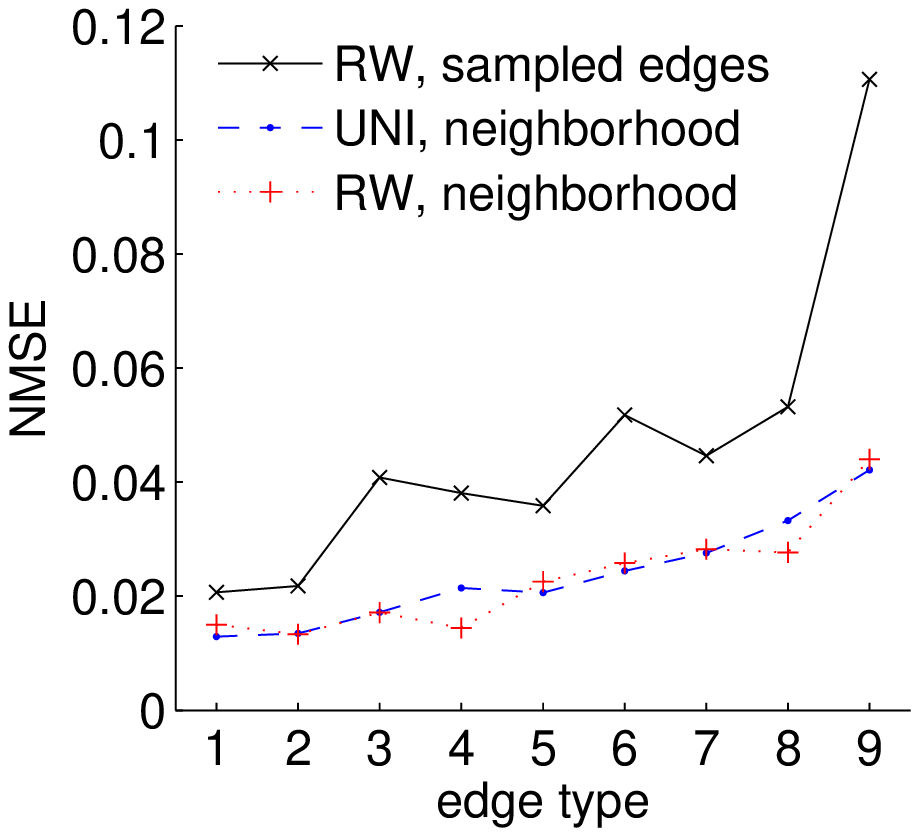}}
\caption{(Xiami) NMSE of edge gender type density estimates.}\label{fig:gendercmp}
\end{figure}

Let us illustrate how to apply the edge label density estimation.
Consider the directed graph of Xiami,
53.8\% of users are male (M), 37.5\% are female (F), and 8.7\% are unknown (U).
A directed edge $(u,v)$ can be classified into the following nine types
when the edge label is defined as $u.gender\rightarrow v.gender$:
1) M$\rightarrow$M, 2) M$\rightarrow$F, 3) M$\rightarrow$U,
4) F$\rightarrow$M, 5) F$\rightarrow$F, 6) F$\rightarrow$U,
7) U$\rightarrow$M, 8) U$\rightarrow$F, 9) U$\rightarrow$U.
Fig.~\ref{fig:genderedgedensity} shows edge density $\btau=(\tau_1,\ldots,\tau_9)$, where $\tau_i$ ($1\le i \le 9$) is the fraction of type $i$ edges.
We can easily find that the fraction of edges with a certain edge type approximately equals the
product of the fractions of nodes with its two endpoints' genders.
This indicates that users' following behaviors in Xiami are not directly related with gender.
Fig.~\ref{fig:gendercmp} shows results for estimating $\btau$. Similarly we can find that RW and UNI using sampled nodes' neighborhood information exhibits small errors, and are two times more accurate than the simple RW method.

\subsection{High Degree Node Detection}
\begin{figure*}[htb]
\center
\subfigure[Xiami]{
\includegraphics[width=0.3\textwidth]{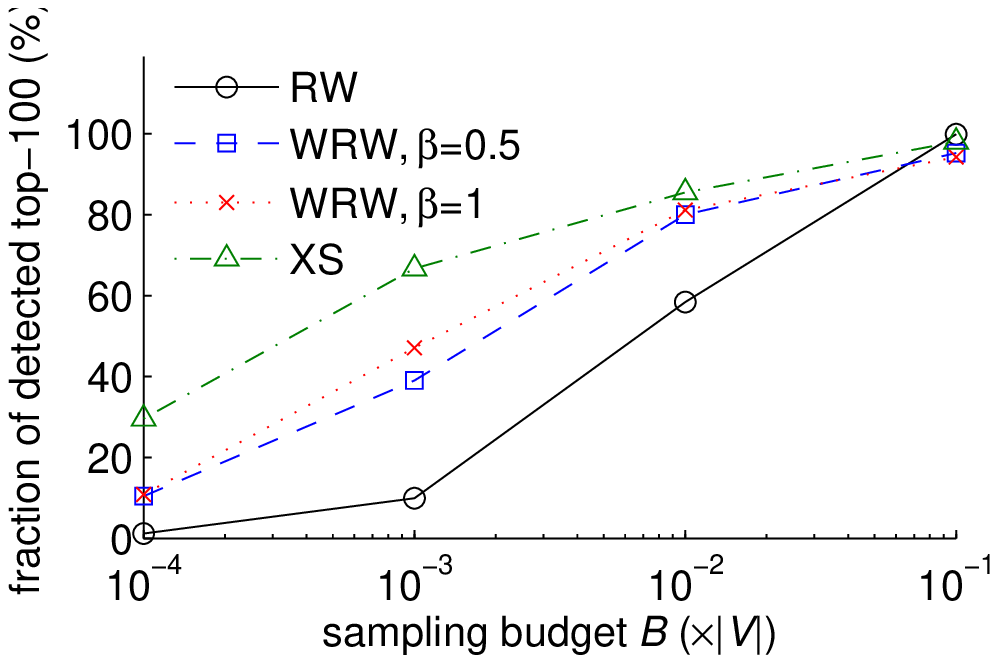}}
\subfigure[Flickr]{
\includegraphics[width=0.3\textwidth]{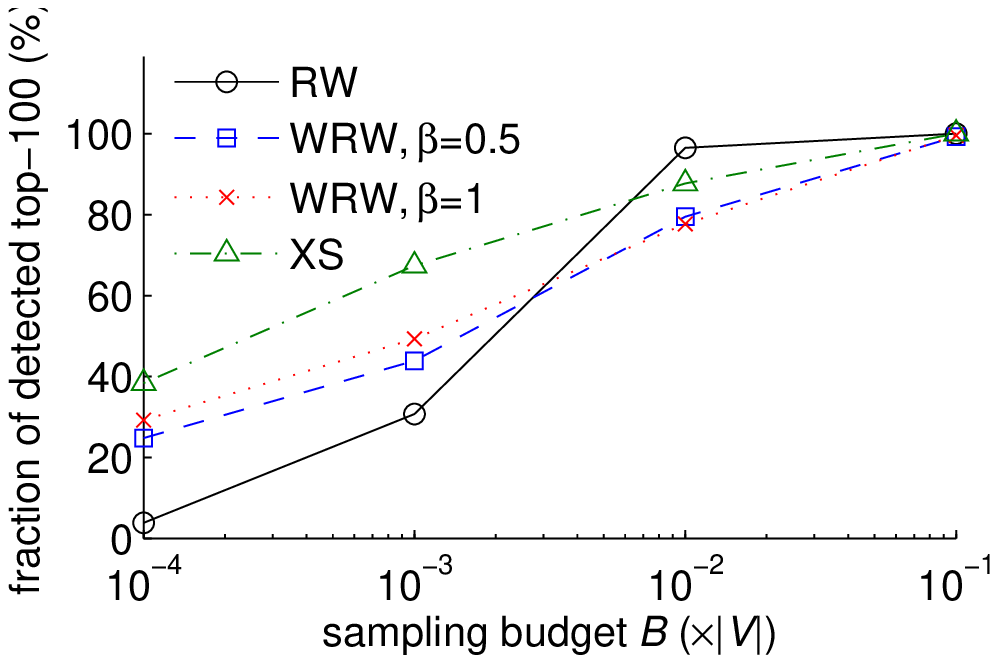}}
\subfigure[YouTube]{
\includegraphics[width=0.3\textwidth]{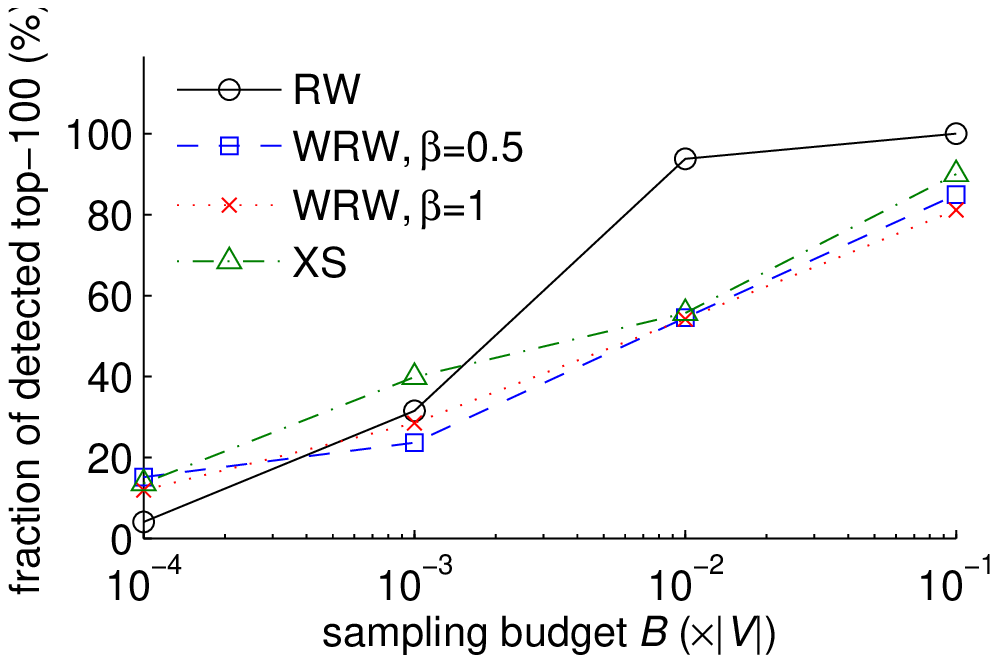}}
\caption{(Previous methods) Results of top-100 high degree node detection.}\label{fig:previoussupernode}
\end{figure*}

\begin{figure*}[htb]
\center
\subfigure[Xiami]{
\includegraphics[width=0.3\textwidth]{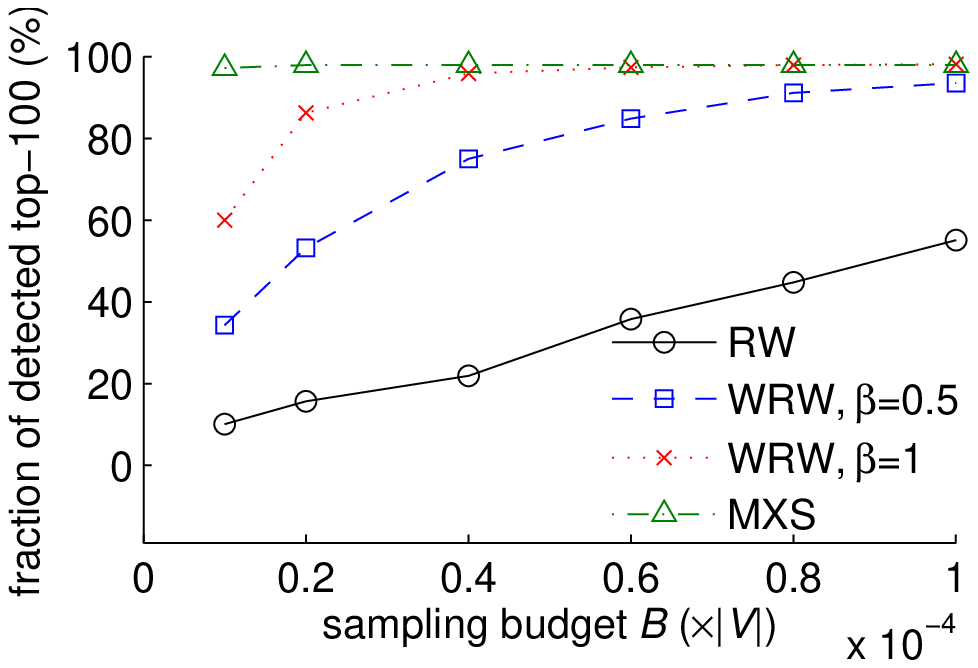}}
\subfigure[Flickr]{
\includegraphics[width=0.3\textwidth]{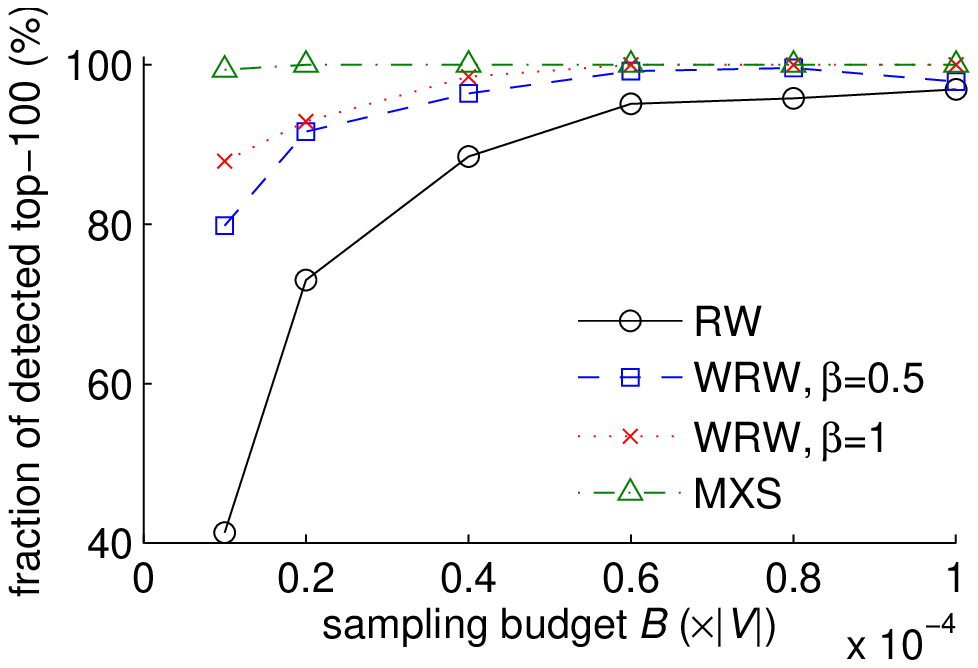}}
\subfigure[YouTube]{
\includegraphics[width=0.3\textwidth]{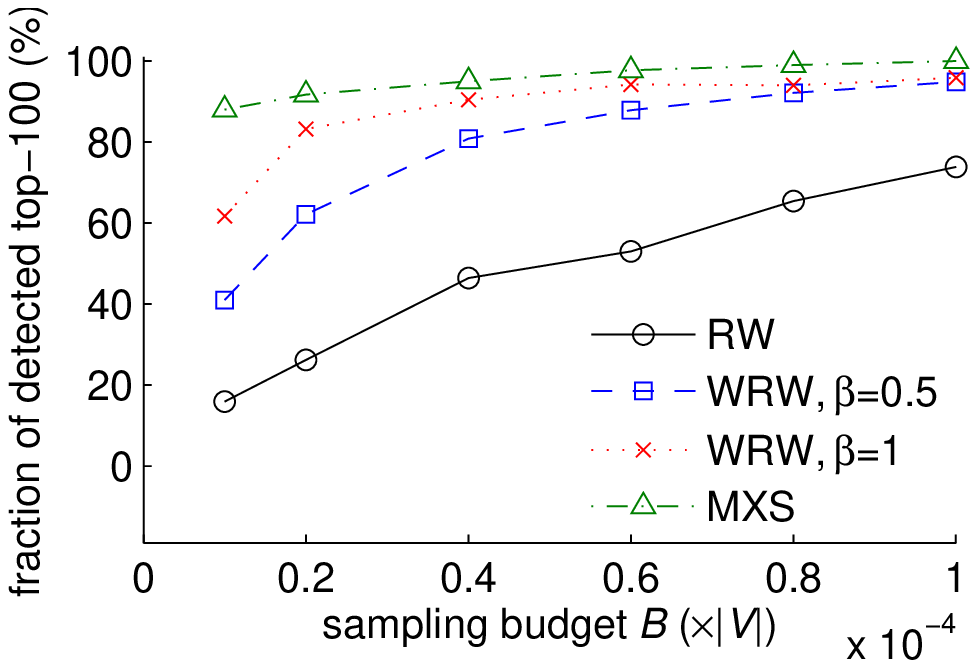}}
\caption{(Our methods, using neighbor information of sampled nodes) Results of top-100 high degree node detection.}\label{fig:supernode}
\end{figure*}

\begin{figure*}[htb]
\center
\subfigure[Xiami]{
\includegraphics[width=0.322\textwidth]{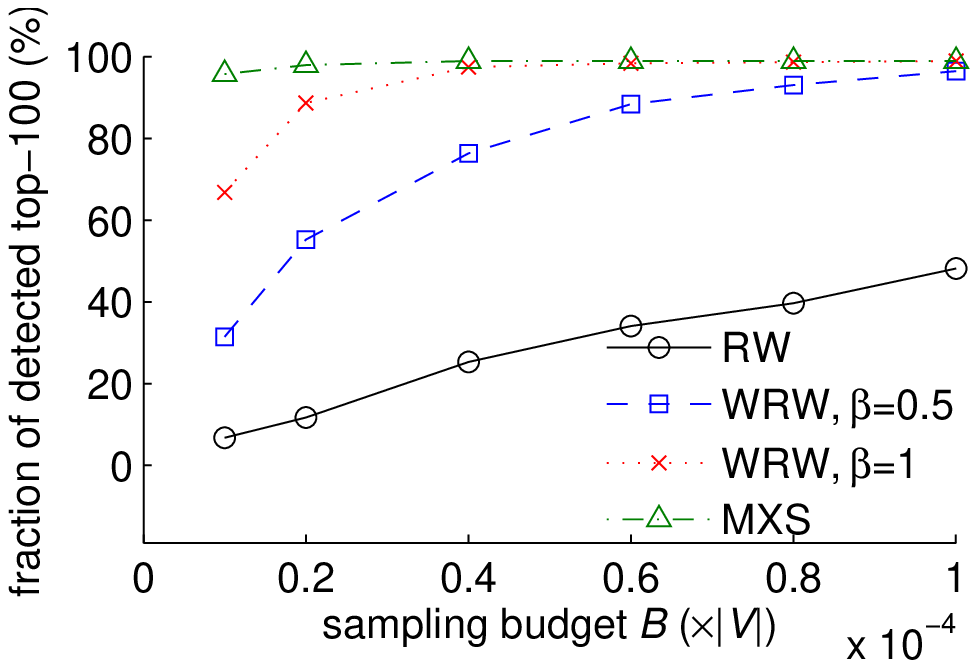}}
\subfigure[Flickr]{
\includegraphics[width=0.322\textwidth]{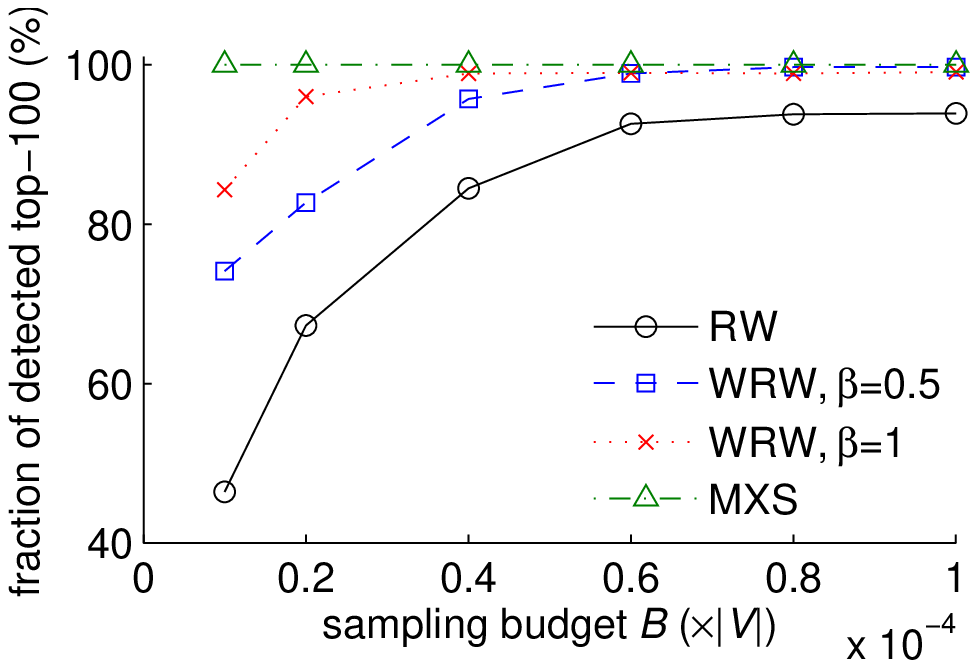}}
\subfigure[YouTube]{
\includegraphics[width=0.322\textwidth]{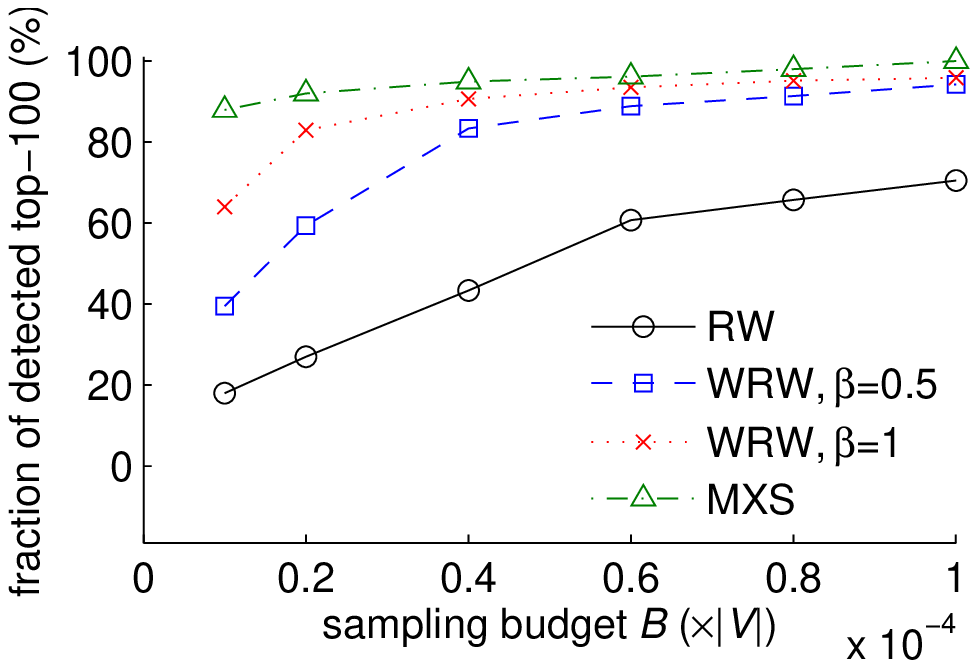}}
\caption{(Our methods, using neighborhood information of sampled nodes) Results of top-100 high out-degree node detection.}\label{fig:supernodeout}
\end{figure*}

\begin{figure*}[htb]
\center
\subfigure[Xiami]{
\includegraphics[width=0.322\textwidth]{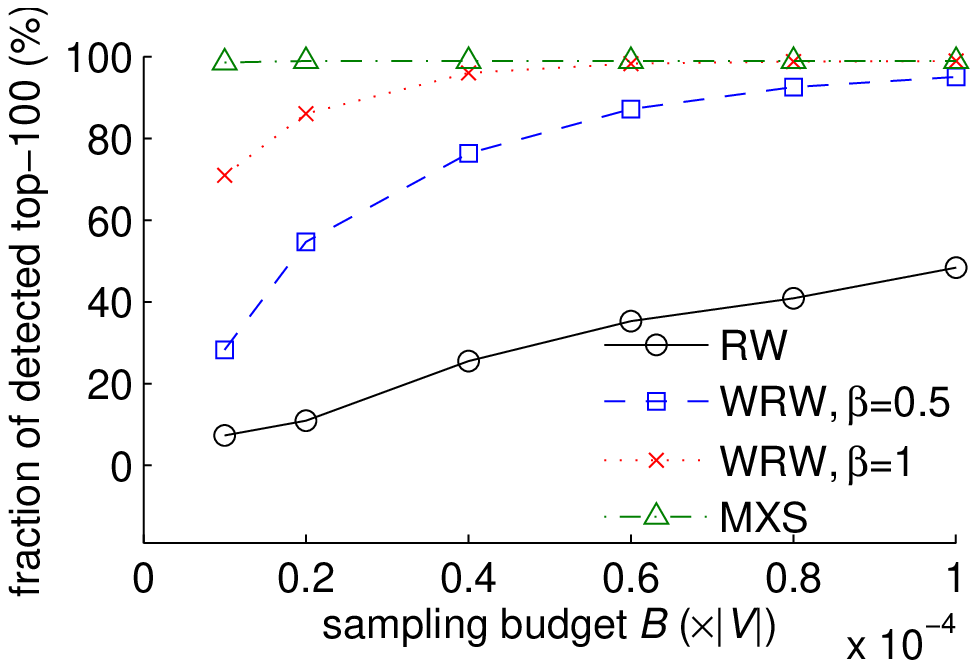}}
\subfigure[Flickr]{
\includegraphics[width=0.322\textwidth]{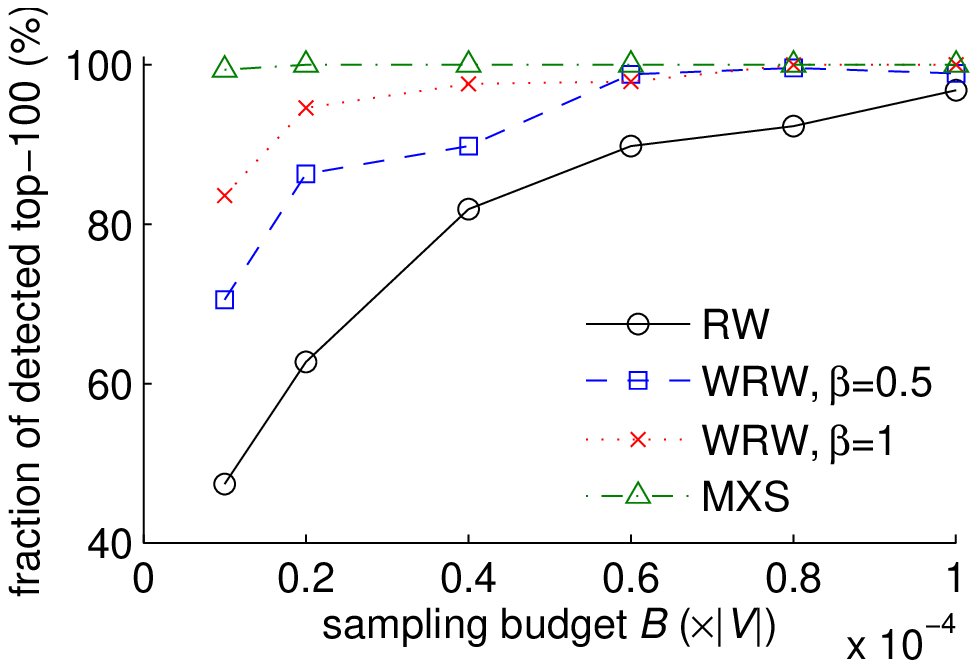}}
\subfigure[YouTube]{
\includegraphics[width=0.322\textwidth]{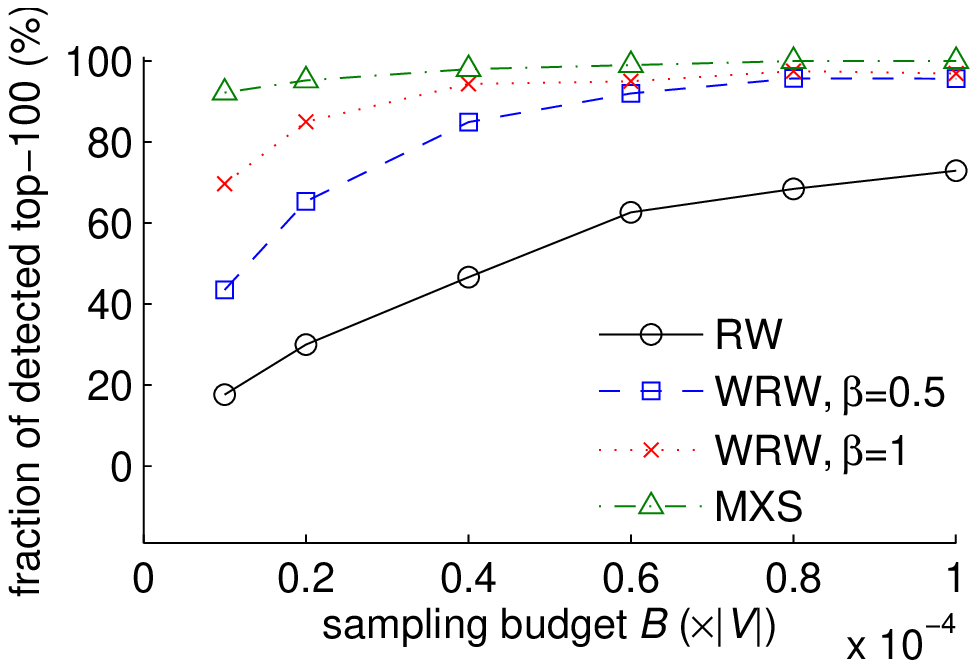}}
\caption{(Our methods, using neighborhood information of sampled nodes) Results of top-100 high in-degree node detection.}\label{fig:supernodein}
\end{figure*}

Fig.\ref{fig:previoussupernode} shows the results of previous methods  for detecting top-100 high degree nodes,
where the edge weight function is defined as $w(u,v)=(d_u d_v)^\beta$ for WRW.
For previous methods without free neighborhood information of sampled nodes, we assume that XS and WRW both must retrieve degree information of a neighbor of sampled nodes with the same cost of sampling a node.
Fig.~\ref{fig:previoussupernode} shows that all of RW, WRW, and XS need to sample more than 10\% of nodes to obtain an accurate result for detecting top-100 degree nodes with the largest degrees.
Fig.~\ref{fig:supernode} shows the results of RW, WRW, and MXS using free neighborhood information of sampled nodes.
A total of 1,000 runs are used to produce the averages seen in the graph.
It shows that RW, WRW, MXS using neighborhood information are much more efficient than previous methods, and MXS outperforms RW and WRW.
We observe that MXS detects almost 90\% of the top-100 high degree nodes based on a very small fraction of sampled nodes, $B=10^{-5}\times|V|$.
Meanwhile, we compare MXS and XS based on the assumption that XS can be implemented at no cost of looking up sampled nodes' neighbor's neighbor information, and find they has little difference. We omit the details here.
Similarly Figs.~\ref{fig:supernodeout} and~\ref{fig:supernodein} show that MXS is much more efficient than the other two methods for detecting top-100 high out-degree nodes and top-100 high in-degree nodes for directed graphs, where each node has the knowledge of its neighbors' out-degrees and in-degrees.
The edge weight function of WRW is defined as $w(u,v)=(d_u^\text{(O)}+d_u^\text{(I)})^\beta (d_v^\text{(O)}+d_v^\text{(I)})^\beta$.

\subsection{Short Path Discovery}
\begin{figure}[htb]
\center
\subfigure[Soc-Epinions]{
\includegraphics[width=0.23\textwidth]{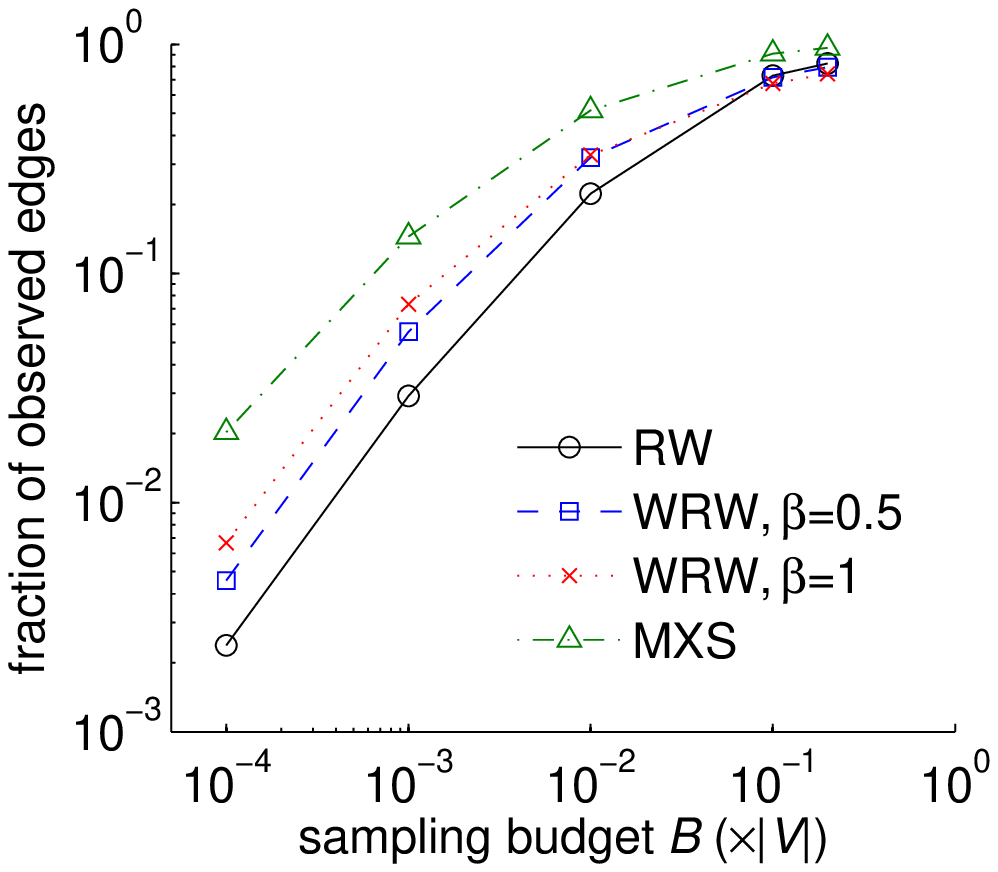}}
\subfigure[Soc-Slashdot]{
\includegraphics[width=0.23\textwidth]{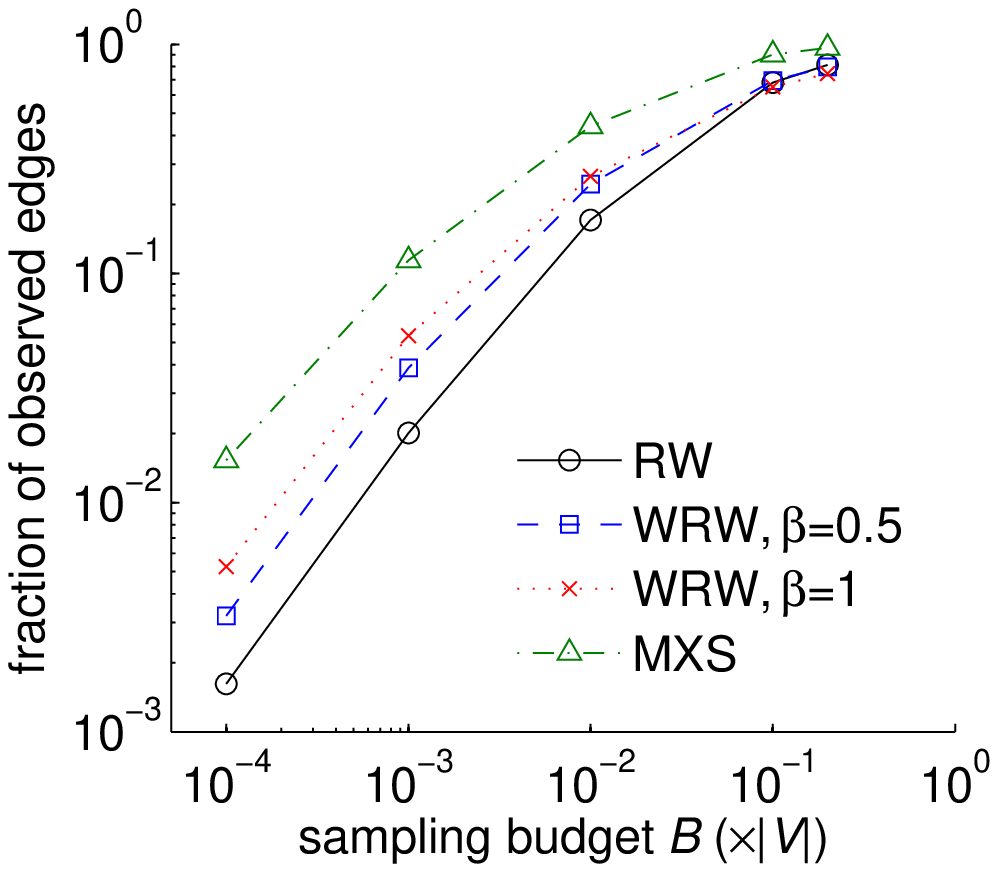}}
\caption{Fractions of observed edges.}\label{fig:edgefraction}
\end{figure}

\begin{figure}[htb]
\begin{center}
\includegraphics[width=0.4\textwidth]{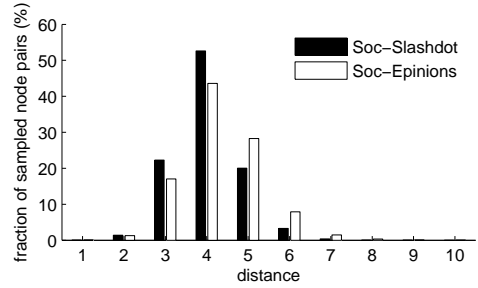}
\caption{(Soc-Slashdot and Soc-Epinions) Fractions of sampled node pairs with given distances.}\label{fig:distancedistribution}
\end{center}
\end{figure}

\begin{figure}[htb]
\center
\subfigure[Soc-Epinions]{
\includegraphics[width=0.23\textwidth]{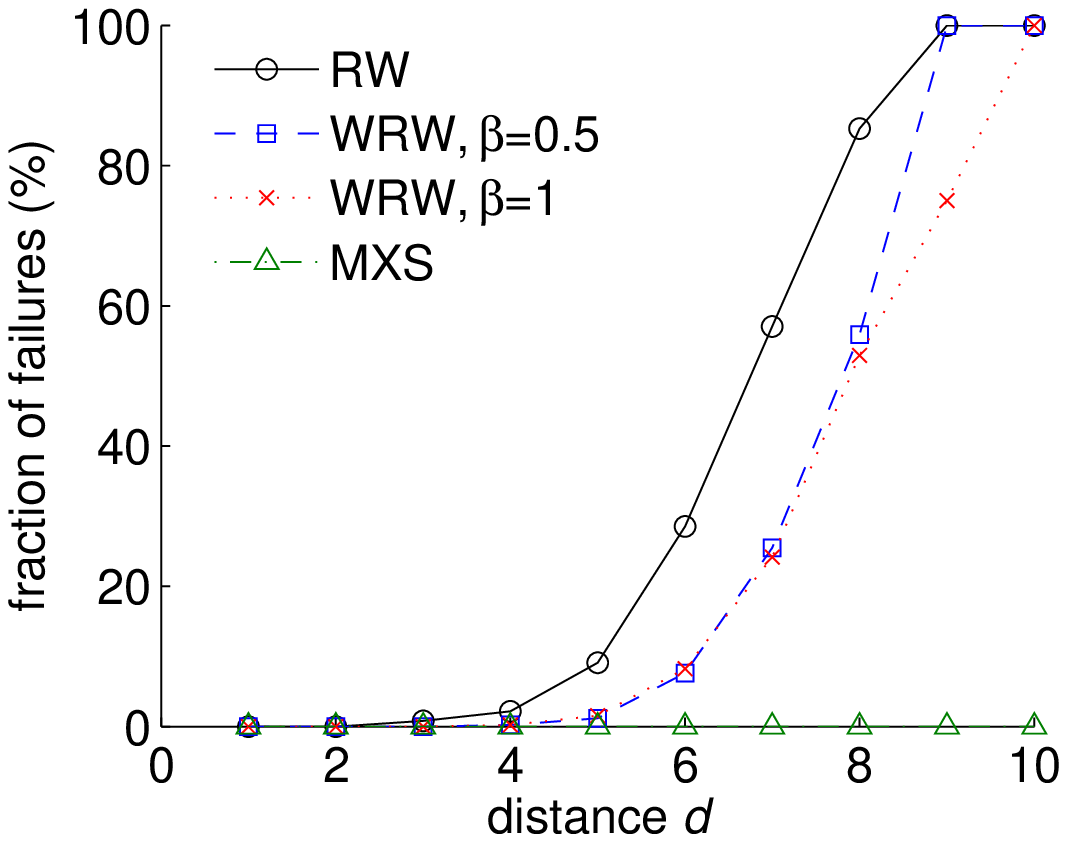}}
\subfigure[Soc-Slashdot]{
\includegraphics[width=0.23\textwidth]{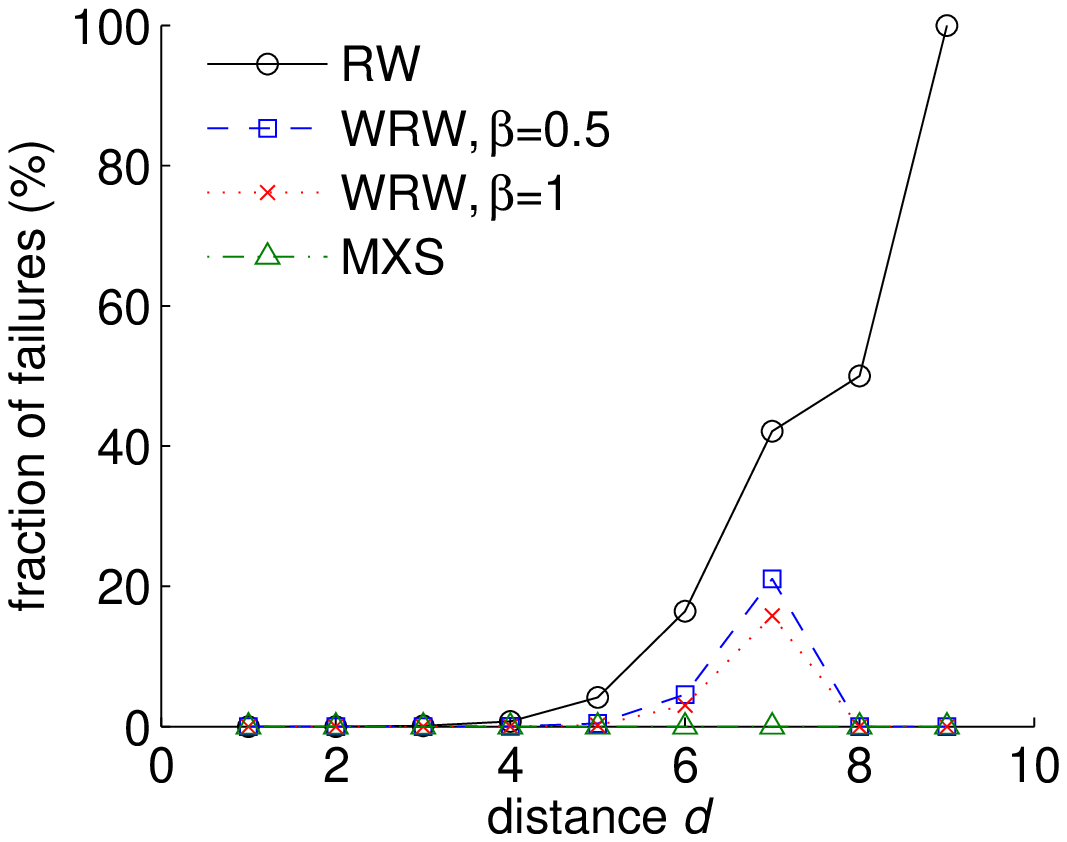}}
\caption{Fractions short path discovery failures, $B=20$.}\label{fig:failureratio}
\end{figure}

\begin{figure}[htb]
\center
\subfigure[Soc-Epinions]{
\includegraphics[width=0.23\textwidth]{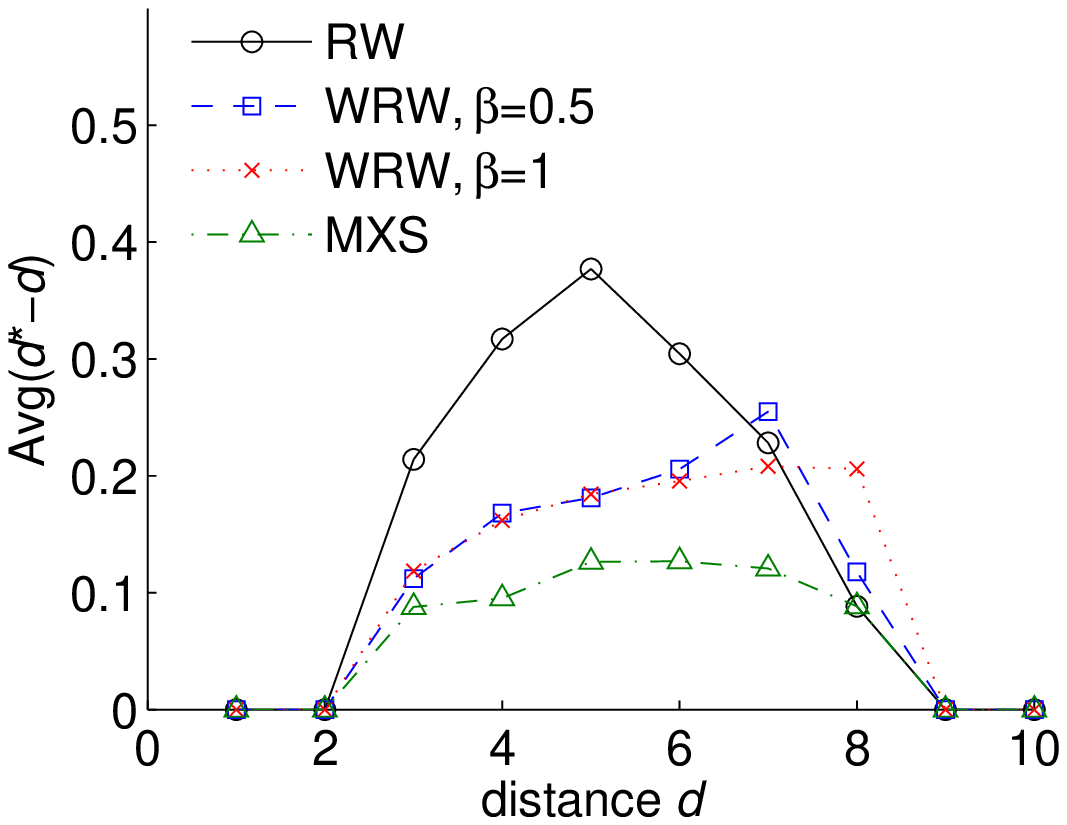}}
\subfigure[Soc-Slashdot]{
\includegraphics[width=0.23\textwidth]{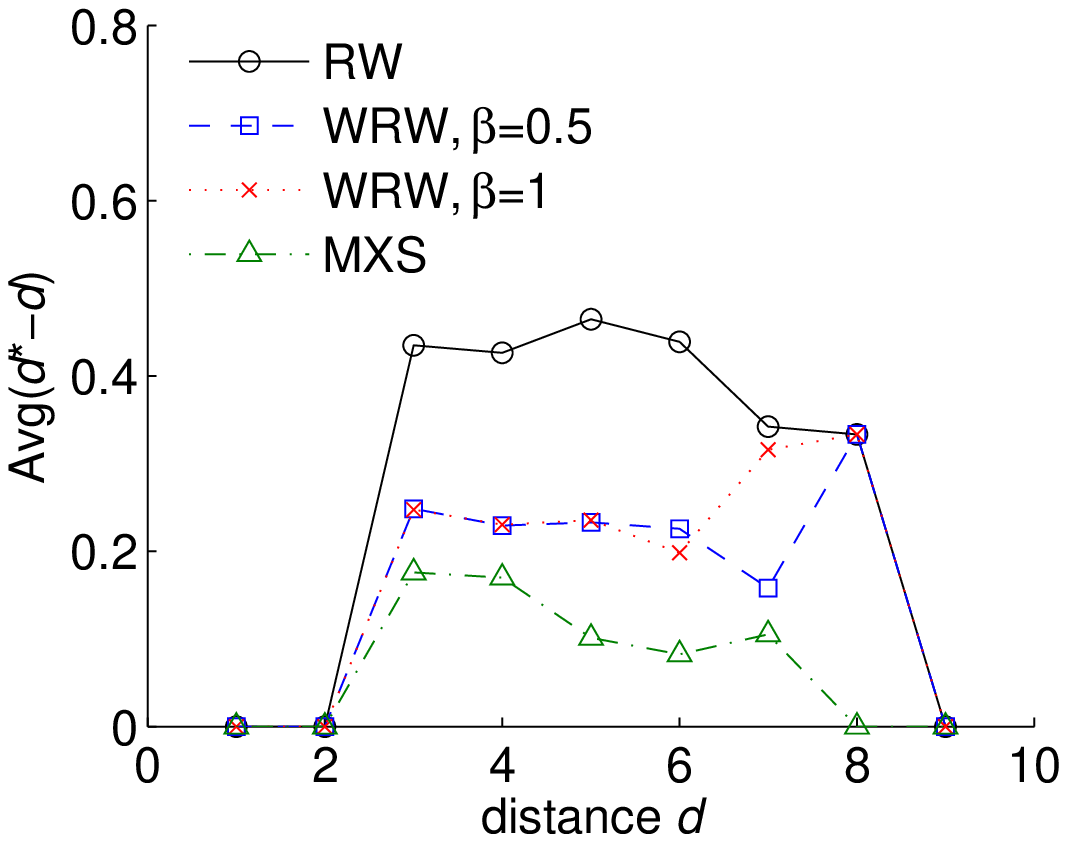}}
\caption{NMSE of short path lengthes, $B=20$.}\label{fig:lengthnmse}
\end{figure}

Fig.~\ref{fig:edgefraction} shows that MXS observes many more edges than WRW and RW under the same sampling budget $B$, where edge weight function is defined as $w(u,v)=(d_u d_v)^\beta$ for WRW. Note that when $\beta=0$, WRW is the same as RW. As $\beta$ increases, we can see that WRW collects more edges.
In what follows we evaluate our MXS and WRW based short path discovery methods compared with the previous RW based method in~\cite{RibeiroNetSci2012}.
For two nodes with distance $d<\infty$ in $G$, let $d^*$ be the length of the short path observed by sampling methods.
When there is no path observed for them, we denote $d^*=\infty$, and a failure is reported.
For all $d^*<\infty$, we use $E[d^*-d]$ as a metric to measure the performance of detecting the shortest paths.
Figs.~\ref{fig:failureratio}-\ref{fig:lengthnmse} show results for 10,000 node pairs generated randomly, where the sampling budget is set as $B=20$. Fig.~\ref{fig:distancedistribution} shows the fractions of sampled node pairs with given distances (length of shortest pathes in original graphes) for Soc-Slashdot and Soc-Epinions.
Fig.~\ref{fig:failureratio} shows the fractions of short path discovery failures as a function of the distance.
Y axis shows the fraction of failures for node pairs with a given distance.
We can see that RW and WRW generate a large number of failures especially for node pairs with a long distance.
However there is almost no failure for our new method MXS.
Moreover Fig.~\ref{fig:lengthnmse} shows that MXS and WRW usually discover shorter paths in comparison with RW.

\section{Applications} \label{sec:application}
\begin{figure}[htb]
\center
\subfigure[Degree distribution]{
\includegraphics[width=0.23\textwidth]{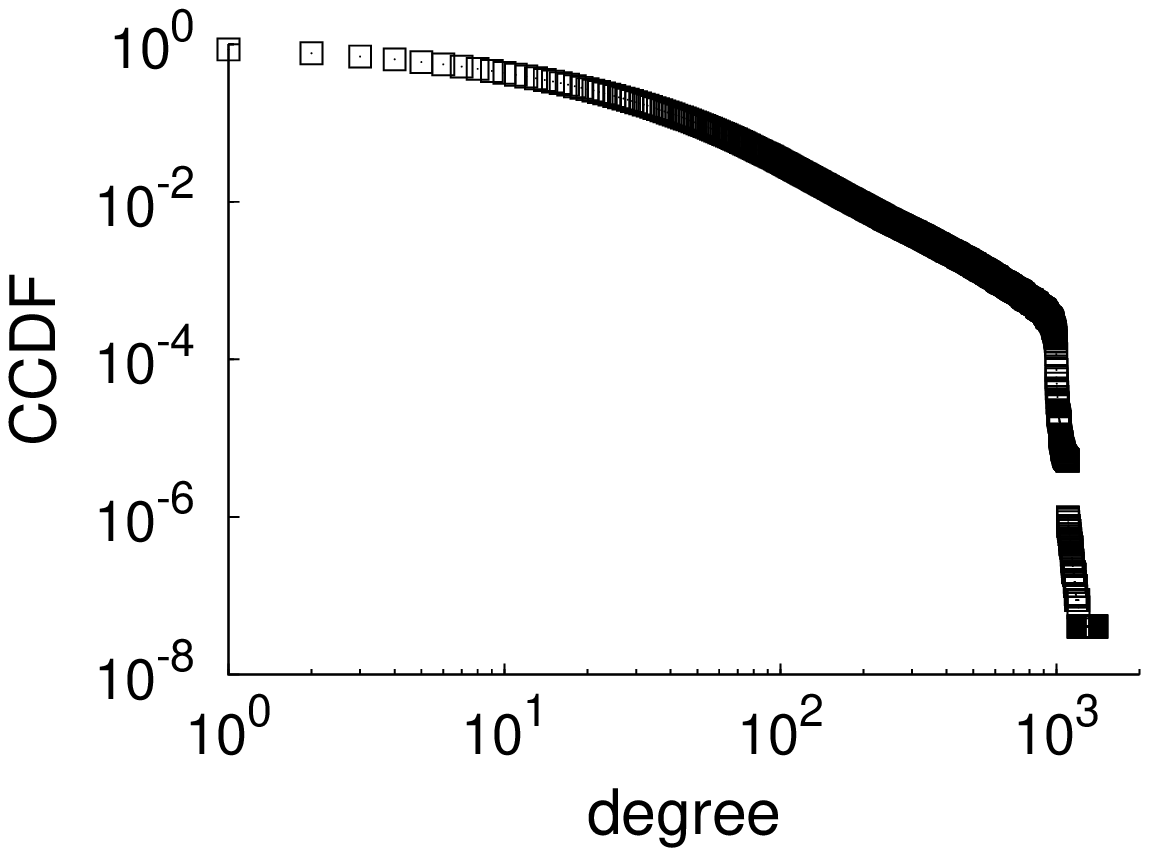}}
\subfigure[Locations distribution]{
\includegraphics[width=0.23\textwidth]{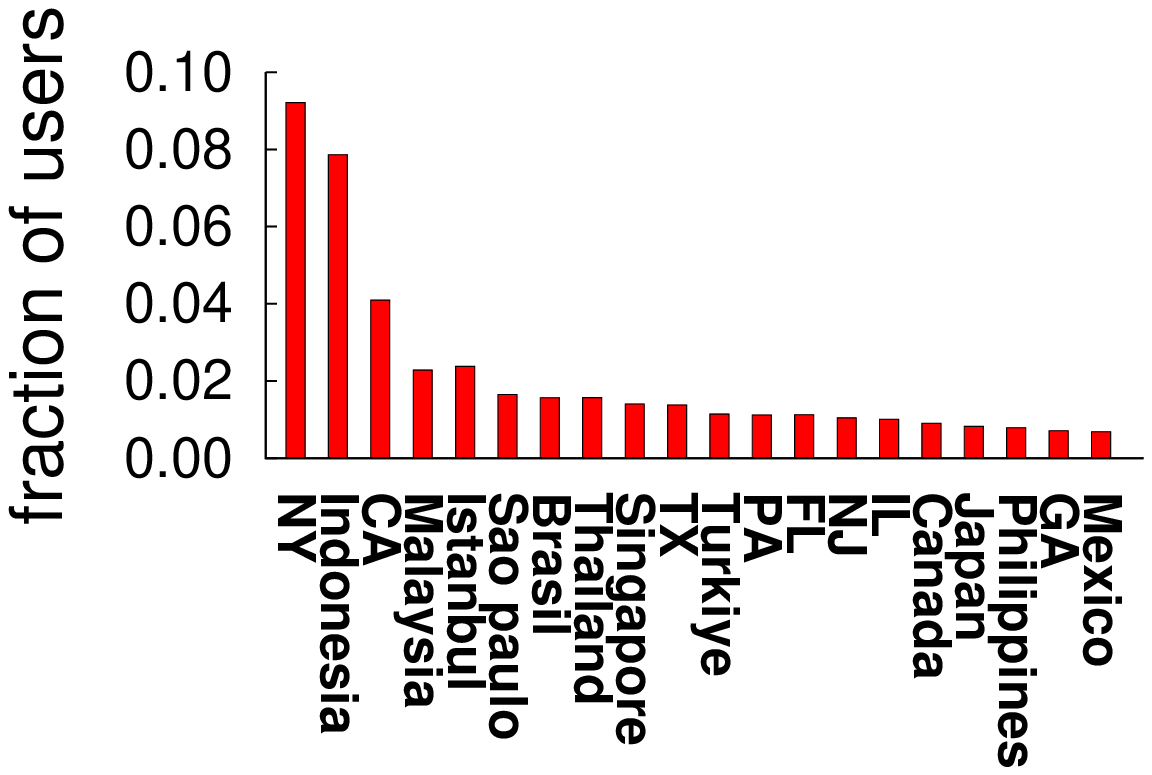}}
\subfigure[Joint degree distribution $\bphi=(\phi(i,j): i\ge j> 0)$]{
\includegraphics[width=0.23\textwidth]{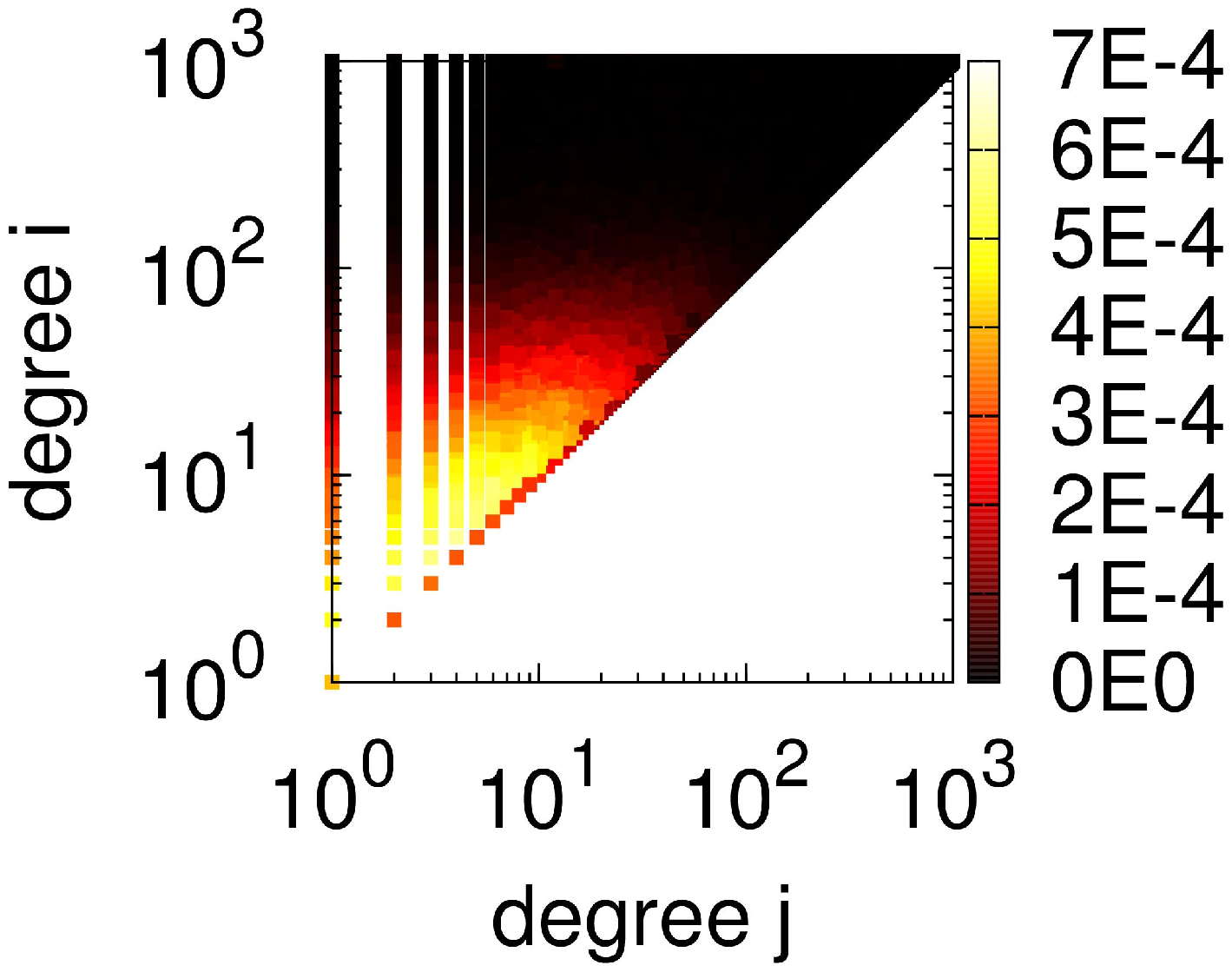}}
\subfigure[High degree node detection]{
\includegraphics[width=0.23\textwidth]{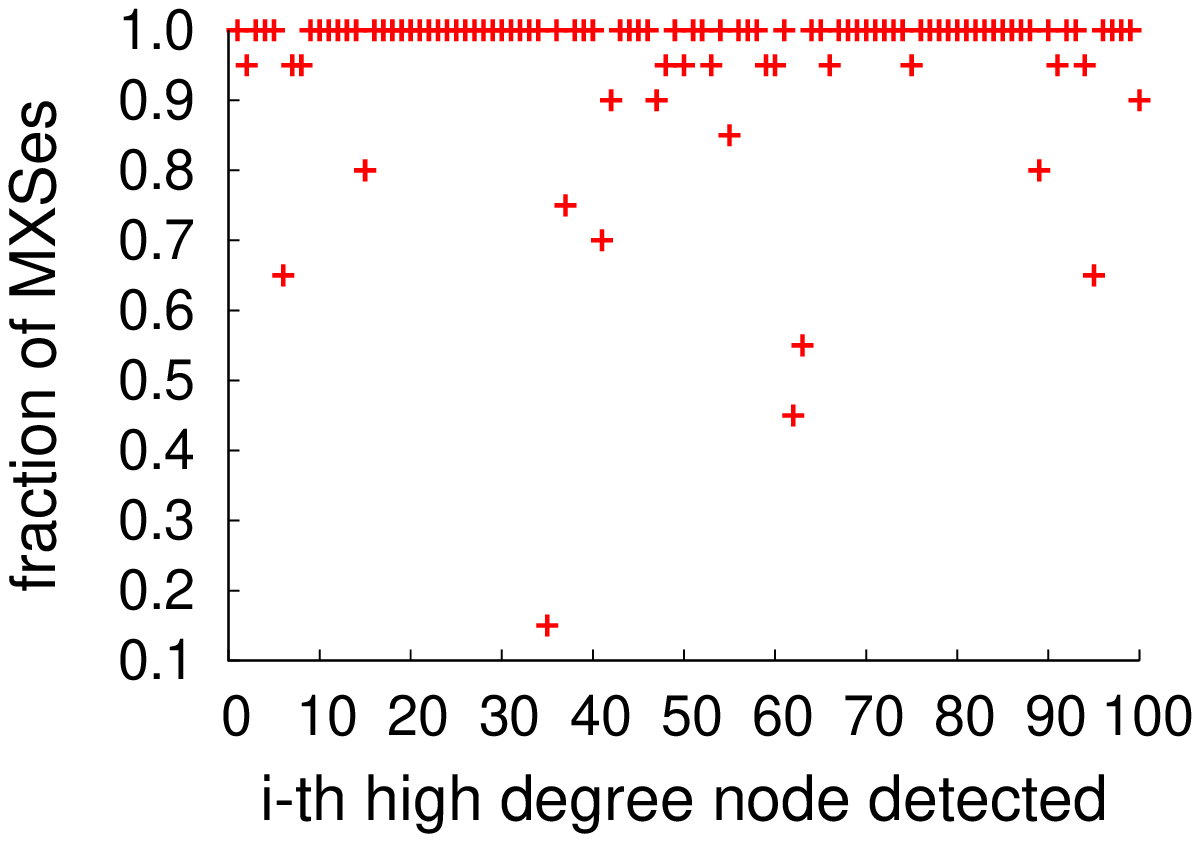}}
\subfigure[Degrees of top-100 high degree nodes]{
\includegraphics[width=0.22\textwidth]{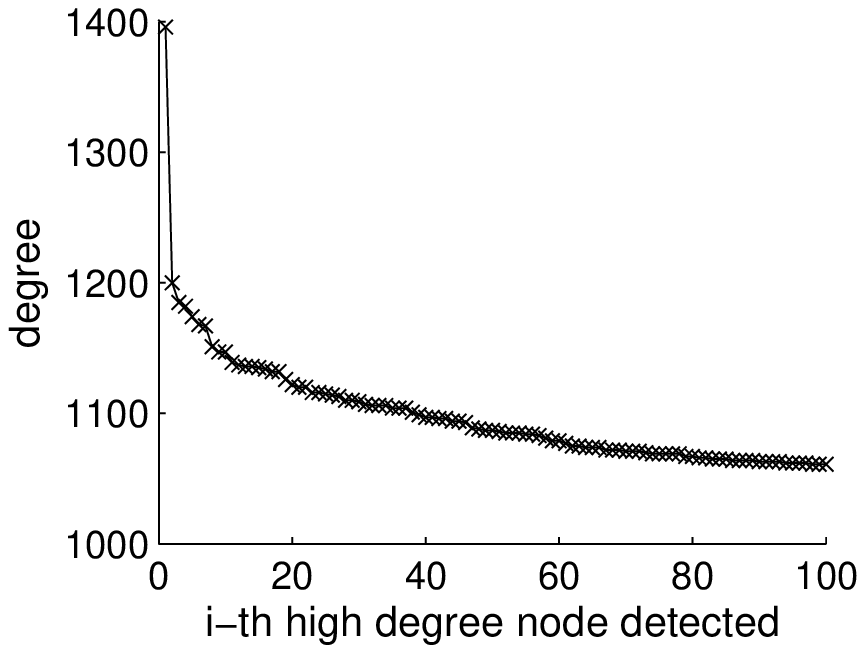}}
\subfigure[Short path discovery]{
\includegraphics[width=0.23\textwidth]{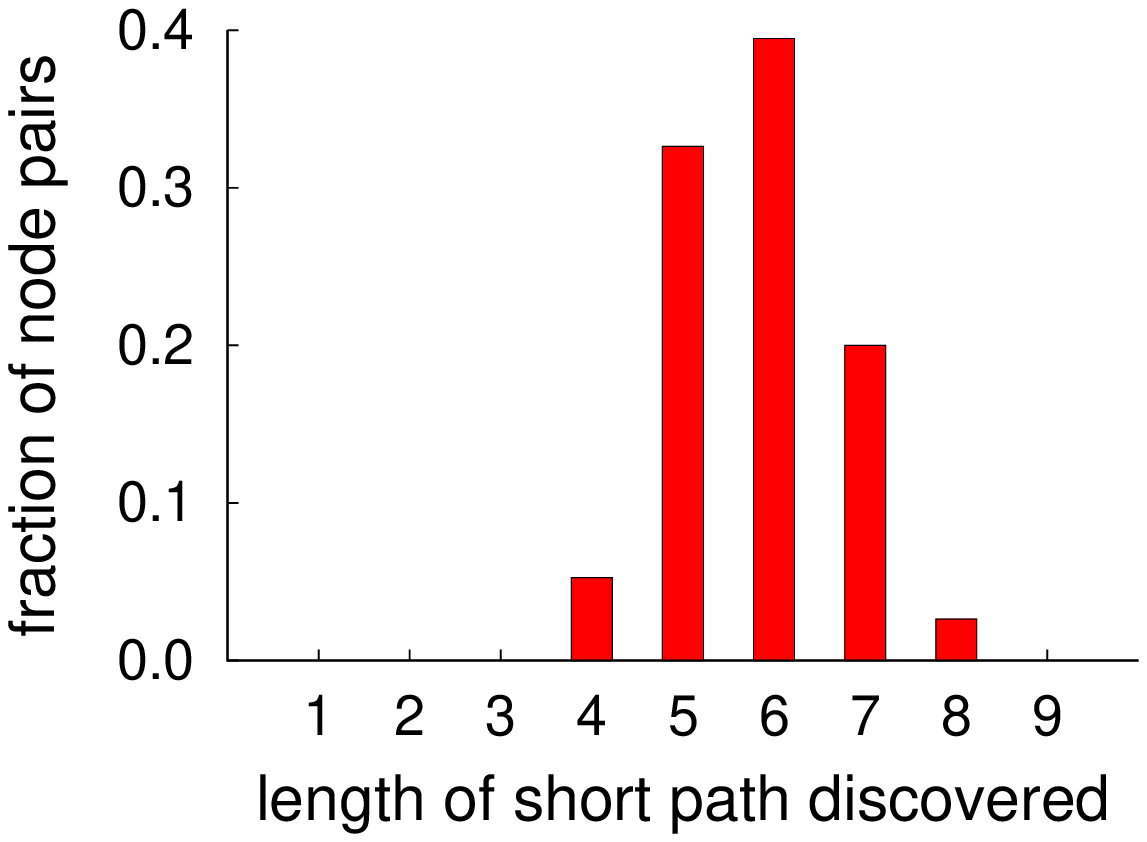}}
\caption{Foursquare structure statistics.}\label{fig:fsdegree}
\end{figure}

Foursquare is a location-based OSN,
which provides web and mobile services for users to explore interested places and leave
tips or comments, and share their check in histories to their friends.
As of December 2012, it has over 25 million active
users\cite{Foursquare_about}.
In Foursquare when we visit a node, we can also
obtain its friends' locations (living places) and degrees.
In what follows, we use our methods which take advantage of
this neighborhood information to characterize the Foursquare graph topology.
Base on $1.9\times 10^5$ users we sampled and their neighborhood information,
we estimate the node degree distribution, joint degree distribution, and location distribution.
To obtain the optimal parameter $\alpha_k$ ($1\le k \le K$) of our mixture estimator in~(\ref{eq:mixestimator}),
we first split sampled nodes into 100 subsets with the same size.
The variance of $\hat\theta_k$ in Equation~(\ref{eq:firstestimatortheta}),
the estimator using sampled nodes,
is computed based on its estimations obtained from 100 node subsets.
Similarly, we estimate the variance of $\breve\theta_k$ in
Equation~(\ref{eq:secondestimatortheta}), the estimator only using neighborhood information.
We then set
$\alpha_k=\frac{\text{Var}(\breve\theta_k)}{\text{Var}(\hat\theta_k)+\text{Var}(\breve\theta_k)}$.
Fig.~\ref{fig:fsdegree} (a) shows results of estimating degree distribution using our mixture estimator based on all sampled nodes.
Average degree is estimated as 21.2.
We can see that the degree distribution of Foursquare {\em does not} exhibit a heavy tail,
which has a sharp drop starting from degree 1,000.
This may be caused by the policy set by Foursqure which limits users
to have only 1,000 friends~\cite{FoursquareLimit}.
From our observed edges, we find that 53.8\% of edges have no node with degree larger than 100.
Fig.~\ref{fig:fsdegree} (c) show results for estimating joint degree distribution $\bphi=(\phi(i,j): i\ge j> 0)$, where $\phi(i,j)$ is the fraction of edges consisting of two nodes with degree $i$ and $j$ separately.
This result is quite interesting for
it shows that friends tend to have similar degrees.
Fig.~\ref{fig:fsdegree} (b) shows Foursquare users' location distribution.
Note that users could provide any text strings to Foursquare as their locations.
This induces different granularity of users' locations.
For example, users from Katsushika-ku in Tokyo, Japan
may reveal their locations as ``Katsushika-ku, Tokyo, Japan'', ``Katsushika-ku, Tokyo'', or ``Tokyo, Japan''.
For simplicity, we split sampled users' location strings by comma, and classify two locations into a same group when they have
at least one similar substring. Hence, there location strings in the above example
will be clustered together and labeled by the most frequency part, say ``Tokyo''.
Due to the limited space, we only show results for top 20 popular locations.
We can see that the most popular location in Foursquare is New York state (NY), which accounts for nearly 10\% of uses. The second popular
location is Indonesia, which accounts for 8\%.

We randomly sample 20 nodes with degrees not smaller than one,
and apply a MXS starting from each random node, where the sampling budget $B$ is 1,000.
For the top 100 nodes with largest degrees in all sampled nodes and their neighbors given by 20 MXSes, we show their frequencies detected by MXSes in Fig.~\ref{fig:fsdegree} (d), where the y-axis is defined as the fraction of MXSes successfully detected the $i$-th high degree node, $1\le i\le 100$.
We can see that most of these high degree nodes are detected by 90\% of MXSes,
which indicates that they may be very close to the ground truth.
These high degree nodes have degrees in the range [1061, 1083],
which are shown in Fig.~\ref{fig:fsdegree}(e).
The top three high degrees are 1396, 1200, and 1185.
Fig.~\ref{fig:fsdegree}(f) shows the length distribution of discovered
short pathes between 20 initially sampled nodes.
Pathes for all 190 node pairs are successfully discovered.
Pathes with lengths 5, 6, and 7 account for 93\%.
The average length is 5.8.

\begin{figure}[htb]
\begin{center}
\includegraphics[width=0.35\textwidth]{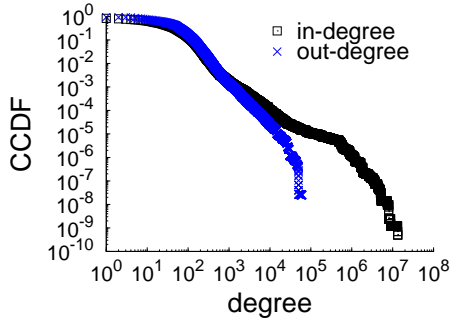}
\caption{(Pinterest) CCDFs of estimated in-degree and out-degree distributions. Estimated average in-degree and out-degree are 60.2 and 61.1, respectively.}\label{fig:pinterest}
\end{center}
\end{figure}

Pinterest is a photo sharing OSN, which allows users to create boards (theme-based image collections) and
pin/repin (collect) images onto their boards from other Pinterest users' image boards and external websites.
Similar to Twitter, Pinterest users can follow other users if they have similar tastes.
In Pinterest when we visit a node we can also obtain its friends' in-degree (number of followers) and out-degree (number of following).
We collected 10,000 users using a RW. The maximum in-degree and out-degree we discovered is 13,331,207, and 61,338. The CCDFs of in-degree and out-degree distributions are shown as Fig.~\ref{fig:pinterest}.

\section{Related Work} \label{sec:related}

Few network sampling methods use neighborhood information to provide accurate estimates that have convergence guarantees.
The work closest to ours is Dasgupta et al.~\cite{Dasgupta2012}.
Dasgupta et al.~ randomly samples nodes (either uniformly or with a known bias) and then uses neighborhood information to improve its unbiased estimator.
However, randomly sampling nodes is practical only if performed uniformly (in our scenarios, rejection sampling to bias the samples makes little sense)
and suffers from low query rate in NoSQL graph databases and Web APIs.
Dasgupta et al.\ partially compensates the low query rate through the use of neighborhood information present in the node query reply
of a number of major OSNs.
Moreover, their estimators require knowledge of degrees of sampled nodes' neighbors, which incurs extra query costs when applied to OSNs such as Pinterest and Sina microblog that do not provide free neighbor degree information.

Kurant et al.~\cite{SIGM_Kurant11} designs a RW-based method that uses a weighted RW to perform stratified sampling on social networks.
These weights are computed using neighborhood information.
Kurant et al.~uses their technique on Facebook and show that their stratified sampling technique achieves higher estimation accuracy than other methods. However, the neighborhood information in their method is limited to helping find random walk weights and not used in the estimator.
Interestingly, our estimator can be easily combined with the weighted random walk in~\cite{SIGM_Kurant11} to improve its accuracy.

Maiya and Berger-Wolf~\cite{KDD_Maiya11} empirically investigates the performance of a number of subgraph sampling methods (e.g., breadth-first search, random walks, etc.) and their performance in respect to various topological properties (e.g., degree, clustering coefficient).
Maiya and Berger-Wolf, however, does not use neighborhood information to improve the estimators or provide convergence guarantees.
The literature also shows a variety of subgraph sampling works without convergence or accuracy guarantees~\cite{Leskovec2006,ICDM_Hubler08},
which have been empirically tested over a variety of networks.
The above works~\cite{KDD_Maiya11,Leskovec2006,ICDM_Hubler08} also consider subgraph sampling techniques that can preserve other metrics, such as the eigenvalues of the original network~\cite{Leskovec2006}, but without accuracy guarantees.

Breadth-First-Search (BFS) introduces a large bias towards high degree nodes, and it is difficult to remove these biases in general, although it can be ameliorated if the network in question is almost random~\cite{KurantJSAC2011}.
Random walk (RW) is  biased to sample high degree nodes, however its bias is known and can be easily corrected~\cite{IMC_Ribeiro10}.
Random walks in the form of Respondent Driven Sampling (RDS)~\cite{Heckathorn2002,Salganik2004} has been used to estimate population densities using snowball samples of sociological studies.
RDS was developed for small social networks with hidden links while our method considers large online social networks without hidden links.

The Metropolis-Hasting RW (MHRW)~\cite{Stutzbach2009} modifies the RW procedure, aimed at sampling nodes with equal probability.
However, in Ribeiro and Towsley~\cite{CDC_Ribeiro12} we prove that MHRW degree distribution estimates perform poorly in comparison to RWs, more markedly for large degree nodes whose error grows proportionally to the degree value.
Empirically, the accuracy of RW and MHRW has beeen compared in~\cite{Rasti2009,Gjoka2010} and, as predicted by our theoretical results, RW is consistently more accurate than MHRW.

\section{Conclusions and Future Work} \label{sec:conclusions}
In this paper, we study the problem of estimating characteristics for
graphs where nodes have knowledge of their neighbors' properties.
This feature is actually quite common in many OSNs, such as Pinterest~\cite{Pinterest},
Foursquare~\cite{Foursquare}, Sina microblog~\cite{SinaMicroblog}, and Xiami~\cite{Xiami}.
We propose efficient network characteristic (degree and edge density distributions) estimators from sampling which have
show provable convergence and accuracy guarantees.
Our method is tailored to NoSQL graph databases (e.g.\ Neo4j) and the type of Web API present in major social network
websites such as Facebook, Google+, Twitter, Pinterest~\cite{Pinterest}, and
Foursquare~\cite{Foursquare}.
We can also adapt known techniques to detect high-degree nodes and short path discovery
between nodes .
Our experimental results show that our estimator drastically reduces (by 4-fold) the number of samples required
to achieve the same estimation accuracy.
Our generalization allows us
to include neighboring information in the estimation of a
variety of network characteristics from nodes sampled
using a random walk-based technique called Frontier Sampling~\cite{IMC_Ribeiro10}.
As future work, we plan to replace Lemma~\ref{lemma:accuracy} with a bound that, for $m>1$, considers
all samples of FS.

\section*{Acknowledgments}
This work was supported by the NSF grant CNS-1065133 and ARL Cooperative Agreement W911NF-09-2-0053. The views and conclusions contained in this document are those of the authors and should not be interpreted as representing the official policies, either expressed or implied of the NSF, ARL, or the U.S. Government. This work was also supported in part by the NSFC funding 60921003 and 863 Program 2012AA011003 of China.

\balance
\bibliographystyle{abbrv}
\bibliography{benefitneighborhood,RibeiroBib}

\begin{thebibliography}{10}

\bibitem{Foursquare_about}
\url{https://foursquare.com/about/}.

\bibitem{Lim2011}
Online estimating the $k$ central nodes of a network.
\newblock In {\em Proceedings of IEEE Network Science Workshop 2011}, June
  2011.

\bibitem{AX_Neville12}
N.~K. Ahmed, J.~Neville, and R.~Kompella.
\newblock Network sampling: From static to streaming graphs.
\newblock {\em arXiv preprint arXiv:1211.3412}, 2012.

\bibitem{AvrachenkovWAW2010}
K.~Avrachenkov, B.~Ribeiro, and D.~Towsley.
\newblock Improving random walk estimation accuracy with uniform restarts.
\newblock In {\em The 7th Workshop on Algorithms and Models for the Web Graph},
  pages 98--109, December 2010.

\bibitem{Cooper2012}
C.~Cooper, T.~Radzik, and Y.~Siantos.
\newblock A fast algorithm to find all high degree vertices in power law
  graphs.
\newblock In {\em Proceedings of WWW 2012 LSNA workshop}, pages 1007--1016,
  April 2012.

\bibitem{Coppersmith1993}
D.~Coppersmith, P.~Doyle, P.~Raghavan, and M.~Snir.
\newblock Random walks on weighted graphs, and applications to on-line
  algorithms (extended).
\newblock {\em Journal of the ACM}, pages 369--378, 1993.

\bibitem{Dasgupta2012}
A.~Dasgupta, R.~Kumar, and D.~Sivakumar.
\newblock Social sampling.
\newblock In {\em Proceedings of ACM SIGKDD 2012}, pages 235--243, August 2012.

\bibitem{Foursquare}
Foursquare.
\newblock \url{http://www.foursquare.com}, 2012.

\bibitem{FoursquareLimit}
Foursquare limit.
\newblock
  \url{http://www.quora.com/Foursquare/What-are-some-of-the-reasons-foursquare%
\\-limits-users-to-having-only-1000-friends}, 2012.

\bibitem{Gjoka2010}
M.~Gjoka, M.~Kurant, C.~T. Butts, and A.~Markopoulou.
\newblock Walking in facebook: A case study of unbiased sampling of {OSNs}.
\newblock In {\em Proceedings of IEEE INFOCOM 2010}, pages 2498--2506, April
  2010.

\bibitem{Heckathorn2002}
D.~D. Heckathorn.
\newblock Respondent-driven sampling {II}: deriving valid population estimates
  from chain-referral samples of hidden populations.
\newblock {\em Social Problems}, 49(1):11--34, 2002.

\bibitem{ICDM_Hubler08}
C.~Hubler, H.-P. Kriegel, K.~Borgwardt, and Z.~Ghahramani.
\newblock Metropolis algorithms for representative subgraph sampling.
\newblock In {\em Data Mining, 2008. ICDM'08. Eighth IEEE International
  Conference on}, pages 283--292. IEEE, 2008.

\bibitem{Hui2008}
P.~Hui, J.~Crowcroft, and E.~Yoneki.
\newblock Bubble rap: Social-based forwarding in delay tolerant networks.
\newblock In {\em Proceedings of ACM MobiHoc 2008}, pages 241--250, May 2008.

\bibitem{SIGM_Kurant11}
M.~Kurant, M.~Gjoka, C.~T. Butts, and A.~Markopoulou.
\newblock Walking on a graph with a magnifying glass: stratified sampling via
  weighted random walks.
\newblock In {\em SIGMETRICS}, volume~39, pages 241--252, 2011.

\bibitem{KurantJSAC2011}
M.~Kurant, A.~Markopoulou, and P.~Thiran.
\newblock Towards unbiased bfs sampling.
\newblock {\em IEEE Journal on Selected Areas in Communications},
  29(9):1799--1809, September 2011.

\bibitem{Kwak2010}
H.~Kwak, C.~Lee, H.~Park, and S.~Moon.
\newblock What is twitter, a social network or a news media?
\newblock In {\em Proceedings of WWW 2010}, pages 591--600, April 2010.

\bibitem{Leskovec2006}
J.~Leskovec and C.~Faloutsos.
\newblock Sampling from large graphs.
\newblock In {\em Proceedings of ACM SIGKDD 2006}, pages 631--636, June 2006.

\bibitem{LeskovecWWW2008}
J.~Leskovec and E.~Horvitz.
\newblock Planetary-scale views on a large instant-messaging network.
\newblock In {\em Proceedings of WWW 2008}, pages 915--924, April 2008.

\bibitem{LeskovecIM2009}
J.~Leskovec, K.~J. Lang, A.~Dasgupta, and M.~W. Mahoney.
\newblock Community structure in large networks: Natural cluster sizes and the
  absence of large well-defined clusters.
\newblock {\em Internet Mathematics}, 6(1):29--123, 2009.

\bibitem{Maiya2010}
A.~S. Maiya and T.~Y. Berger-Wolf.
\newblock Online sampling of high centrality individuals in social networks.
\newblock In {\em Proceedings of the 14th Pacific-Asia Conference on Knowledge
  Discovery and Data Mining (PAKDD2010)}, pages 91--98, June 2010.

\bibitem{KDD_Maiya11}
A.~S. Maiya and T.~Y. Berger-Wolf.
\newblock Benefits of bias: towards better characterization of network
  sampling.
\newblock In {\em SIGKDD}, pages 105--113, 2011.

\bibitem{MisloveIMC2007}
A.~Mislove, M.~Marcon, K.~P. Gummadi, P.~Druschel, and B.~Bhattacharjee.
\newblock Measurement and analysis of online social networks.
\newblock In {\em Proceedings of ACM SIGCOMM Internet Measurement Conference
  2007}, pages 29--42, October 2007.

\bibitem{PageRank}
L.~Page, S.~Brin, R.~Motwani, and T.~Winograd.
\newblock The pagerank citation ranking: bringing order to the web.
\newblock Technical report, Stanford InfoLab, 1999.

\bibitem{Pinterest}
Pinterest.
\newblock \url{http://www.pinterest.com}, 2013.

\bibitem{Rasti2009}
A.~H. Rasti, M.~Torkjazi, R.~Rejaie, N.~Duffield, W.~Willinger, and
  D.~Stutzbach.
\newblock Respondent-driven sampling for characterizing unstructured overlays.
\newblock In {\em Proceedings of IEEE INFOCOM Mini-conference 2009}, April
  2009.

\bibitem{RibeiroNetSci2012}
B.~Ribeiro, P.~Bash, and D.~Towsley.
\newblock Multiple random walks to uncover short paths in power law networks.
\newblock In {\em Proceedings of IEEE Infocom NetSciCom Workshop 2012}, pages
  1--6, April 2012.

\bibitem{IMC_Ribeiro10}
B.~Ribeiro and D.~Towsley.
\newblock Estimating and sampling graphs with multidimensional random walks.
\newblock In {\em Proceedings of ACM SIGCOMM Internet Measurement Conference
  2010}, pages 390--403, November 2010.

\bibitem{CDC_Ribeiro12}
B.~Ribeiro and D.~Towsley.
\newblock On the estimation accuracy of degree distributions from graph
  sampling.
\newblock In {\em IEEE Conference on Decision and Control}, 2012.

\bibitem{INFOCOM_Ribeiro2012}
B.~Ribeiro, P.~Wang, F.~Murai, and D.~Towsley.
\newblock Sampling directed graphs with random walks.
\newblock In {\em Proceedings of IEEE INFOCOM 2012}, pages 1692--1700, April
  2012.

\bibitem{Richardson2003}
M.~Richardson, R.~Agrawal, and P.~Domingos.
\newblock Trust management for the semantic web.
\newblock In {\em Proceedings of the 2nd International Semantic Web
  Conference}, pages 351--368, October 2003.

\bibitem{Salganik2004}
M.~J. Salganik and D.~D. Heckathorn.
\newblock Sampling and estimation in hidden populations using respondent-driven
  sampling.
\newblock {\em Sociological Methodology}, 34:193--239, 2004.

\bibitem{SinaMicroblog}
Sina microblog.
\newblock \url{http://weibo.com}, 2012.

\bibitem{Stutzbach2009}
D.~Stutzbach, R.~Rejaie, N.~Duffield, S.~Sen, and W.~Willinger.
\newblock On unbiased sampling for unstructured peer-to-peer networks.
\newblock {\em IEEE/ACM Transactions on Networking}, 17(2):377--390, April
  2009.

\bibitem{PinghuiContent2012}
P.~Wang, J.~Zhao, X.~Guan, J.~C. Lui, and D.~Towsley.
\newblock Sampling contents distributed over graphs.
\newblock Technical Report TR-1201, Xi'an Jiaotong University, 2012.

\bibitem{Xiami}
Xiami.
\newblock \url{http://www.xiami.com}, 2012.

\end{thebibliography}
\end{document}